\documentclass
[showpacs,twocolumn,prb,eqsecnum,hyperref=pdftex,amsmath,longbibliography]{revtex4-1}
\usepackage{hyperref}
\usepackage{comment}
\usepackage{amsmath}
\usepackage{epsfig,color}
\usepackage{graphicx}
\usepackage{dcolumn}
\usepackage{bm}
\usepackage{lineno}
\hypersetup{colorlinks=true,linkcolor=blue,anchorcolor=blue,citecolor=blue,filecolor=blue,urlcolor=blue,bookmarksnumbered=true,pdfview=FitB}
\newcommand{\nn}{\nonumber}
\newcommand{\beq}{\begin{equation}}
\newcommand{\eeq}{\end{equation}}
\newcommand{\bea}{\begin{eqnarray}}
\newcommand{\eea}{\end{eqnarray}}
\newcommand{\bwt}{\begin{widetext}}
\newcommand{\ewt}{\end{widetext}}

\def\k{{\bf k}}
\def\p{{\bf p}}
\def\q{{\bf q}}

\def\r{{\hat \rho}}
\def\cs{^{c(s)}}
\def\nn{\nonumber}
\def\pd{\partial}
\begin{document}

\title{
Fermi-liquid theory and
Pomeranchuk instabilities:\\
fundamentals and new developments }
\author{Andrey V. Chubukov$^{1}$, Avraham Klein$^1$, and Dmitrii L. Maslov$^{2}$}
\affiliation{
$^{1}$Department of Physics, University of Minnesota, 116 Church Street, Minneapolis, MN 55455\\
$^{2}$Department of Physics, University of Florida, P. O. Box 118440, Gainesville, FL
32611-8440
}

\date{\today}

\begin{abstract}
This paper is a short review on the foundations and recent advances in the microscopic Fermi-liquid (FL) theory. We demonstrate that this theory is built on five identities, which follow from conservation of total charge (particle number), spin, and momentum in a translationally and $SU(2)$-invariant FL. These identities allows one to express the effective mass and quasiparticle residue in terms of an exact vertex function and also impose constraints
on the ``quasiparticle'' and ''incoherent" (or ``low-energy'' and ``high-energy'') contributions to the observable quantities. Such constraints forbid certain Pomeranchuk instabilities of a FL,
e.g., towards phases with order parameters that coincide with charge and spin currents.
We provide diagrammatic derivations of these constraints and of the general (Leggett) formula
for the susceptibility in arbitrary angular momentum channel, and illustrate the general relations through simple examples treated in the perturbation theory.
\end{abstract}
\maketitle
It is our great pleasure to write this article for the special volume in celebration of the 85th birthday of Lev Petrovich Pitaevskii,  who, in our view, is one the greatest physicist of his generation and the model of a scientist and
 citizen.  His seminal
volumes on Statistical Physics (with E. M. Lifshitz),\cite{Lifshitz1980}
 Physical Kinetics\cite{physkin} (also with E. M. Lifshitz),  and Quantum Electrodynamics\cite{Landau:QED} (with V. B. Berestetskii and E. M. Lifshitz), all parts of
the Landau and Lifshitz Course on Theoretical Physics,
as well as on Bose-Einstein condensation and superfluidity\cite{BEC} (with S. Stringari)
are not only
 used by every
 contemporary
 physicist
 but
 also,
as we are positive,  will
 serve
 the
 future generations of scientists from around the world.

The paper we present for this volume is
devoted to the
  microscopic theory of a Fermi liquid,
  which was pioneered by Lev Petrovich in the early 1960s.
  The general
  relations he obtained with Landau,  which express the effective mass $m^*/m$ and the quasiparticle residue $Z$ in terms of the vertex function (the Pitaevskii-Landau  relations),  set the gold
   standard for
  many-body theory.
    We hope that  Lev Petrovich and other readers of this volume will find our summary of recent developments in this field interesting.

\section{Introduction}

Despite its apparent simplicity, the
 Fermi-liquid
  (FL) theory is one of
the most non-trivial theories of interacting fermions.~
\cite{Landau0,Landau1,Landau2,agd,Lifshitz1980,Nozieres1999,baym:book,Anderson1984}  In general terms, it states that a
  system of interacting fermions in dimension $D >1$ displays
behavior which differs from that of
 free fermions
 quantitatively rather than qualitatively.
   In particular,
   the FL theory
    states that at temperatures much lower than the Fermi energy $E_F$,
the inverse lifetime of
a fermionic
state near the  Fermi surface (FS) is
 much smaller than
 its
 energy, so that to first approximation these states can be viewed as
  sharp energy levels with energy $\epsilon_\p =
 v_F (p-p_F)$,
measured from $E_F$.
 The only two differences  with free fermions are i) the velocity of excitations $v_F$ is replaced by the effective Fermi velocity $v^*_F$ (or, equivalently, the fermionic mass $m= p_F/v_F$ is replaced by  the effective mass $m^* = p_F/v_F^*$) and ii) the wave function
of a state
 near the FS
 in an interacting system
is renormalized by a factor of $\sqrt{Z} <1$,
 so that the
corresponding probability for the state to be occupied
 is renormalized by $Z$. The factor $Z$ is often called quasiparticle residue. Its presence reflects the fact that even infinitesimally close
to the FS,
 the spectral function of interacting fermions is not just a $\delta$-function, like  for free fermions, but
 also contains
 incoherent background, which extends to energies both above and below $E_F$.
 The
fact that the residue $Z$
 of the $\delta-$functional piece is less than unity implies that interactions moves
 some spectral weight into incoherent background.

It is customary to consider two groups of fermionic states: near the FS and away from it.
We will be referring to the first group as to ``high-energy fermions'', or simply as to ``high energies",
and to the second one as to ``low-energy fermions'', or simply to ``low energies".
 The conventional wisdom is that
 the
 fundamental properties of a FL,
 such as
  its thermodynamic characteristics at low $T$, are completely determined by
 low-energy fermions
 while  high-energy fermions
 can be safely integrated out,
 e.g.,
 within
 the
 renormalization group formalism.~\cite{Shankar1994}
 On a technical level,
high-energy fermions
 are believed
 to determine only the value of
  $Z$ and the vertex function $\Gamma^\omega (p_F, q_F)$,
  which parametrizes
   the interaction between
  low-energy fermions.

 The potential instabilities of a FL -- superconductivity
  and a spontaneous deformation of
  the FS (a Pomeranchuk
  instability)\cite{Pomeranchuk}--
     are also believed to be fully determined by the interaction between
    low-energy
    fermions.
    In particular, the condition  for a Pomeranchuk instability to occur
     in
    the
      charge or spin  channel with orbital momentum $l$ is
      given by $F^{c(s)}_l =-1$, where $F^{c(s)}_l$ are
      the
      Landau parameters, which are
      partial components of
      properly normalized
      $\Gamma^\omega (p_F,q_F)$

Such wisdom, however, is based on the phenomenological formulation of the FL theory,
originally developed by
 Landau\cite{Landau0,Landau1}. In this formulation, one deals exclusively
with low-energy fermions. On the contrast, the microscopic theory of a FL, developed later by
 Landau,\cite{Landau2} Pitaevskii \cite{Pitaevskii1960} and others,\cite{Lifshitz1980,baym:book} allows one to express the fundamental properties
of a FL, such as $m^*/m$, $Z$, and charge and spin susceptibilities, in terms of the exact vertices parameterizing the interactions between {\it all} states. Depending on the particular realization of a FL, as well as on the property considered, the result may or may not be expressed solely via low-energy fermions. One example is the effective mass,
which happens to be a low-energy property only for a Galilean-invariant, or, more generally, Lorentz-invariant FL,\cite{baym:book,baym:1976}
but contains a high-energy contribution otherwise.

The interest to microscopic foundations of the FL theory has intensified over the last few decades due to ubiquitous observations of non-FL behaviors in a wide variety of solid-state systems, such as the cuprate and Fe-based high-temperature superconductors, bad metals,
and other itinerant-electron systems driven to the vicinity of a quantum phase transition.
  The hope is  that if we understand
better the conditions for the FL theory to work, we
will gain a better insight into its failures in these and other cases. Another stimulus for such interest is that currently there are several real-life examples of electronic nematic order, which sets in as a result of a Pomeranchuk instability, e.g., quantum Hall systems,\cite{Fradkin2010} Sr$_3$Ru$_2$O$_7$,\cite{Fradkin2010}  and Fe-based superconductors.~\cite{Fernandes2017}
  The theoretical literature is abundant with proposals for even more esoteric nematic states,
and it is important to understand which of them are feasible.

  In this communication we review earlier and recent work on the microscopic theory of a FL,
  with special attention paid
  to
 the interplay between contributions from
 high- and low-energy
 fermions.
 Our central message is that conservation laws set up delicate balances between these contributions, with sometimes surprising effects, and it is not always possible to
  reduce the high-energy
 contributions
 to
 mere
 renormalizations of
 the
 input parameters for the
 low-energy theory.  The  most spectacular example
  of this
   are the susceptibilities
 of
  the
 charge-current
 and
  spin-current order parameters:
 $\boldsymbol{\r}^c_{J}  (\q) = \sum_{\p,\alpha} \frac{\partial \epsilon_\p}{\partial \p}c^{\dagger}_{\p-\q/2,\alpha}
  c^{\phantom{\dagger}}_{\p+\q/2,\alpha}$
  and
   $\boldsymbol{\r}^s_{J} (\q) = \sum_{\p,\alpha\beta} \frac{\partial \epsilon_\p}{\partial \p}c^{\dagger}_{\p-\q/2,\alpha}
 \sigma^z_{\alpha\beta}
  c_{\p+\q/2,\beta}^{\phantom{\dagger}}$. Within the random phase approximation (RPA), which includes only the low-energy contributions,
  both susceptibilities
  behave
   as
  $\chi^{\cs}_J \propto (m^*/m)/(1 + F^{c(s)}_{1})$ and diverge
  at
    $F^{c(s)}_{1} =-1$, as is expected
  within the
  Pomeranchuk
  scenario.\cite{Pomeranchuk}
However, when
one includes both
   high- and low-energy contributions
  and utilizes the continuity equation associated with
  conservation of
  total charge (particle number) or total spin,
  one
  finds
   that  the divergent piece in  $\chi^{\cs}_J$  cancels out, and   $\chi^{\cs}_J$ remains finite
  at $F^{\cs}_{1} =-1$.
  As
  a consequence,
  a
   Pomeranchuk instability towards the phase with an order parameter
   $\boldsymbol{\r}^c_{J}  (\q)$ or $\boldsymbol{\r}^s_{J}  (\q)$
   cannot occur.
 The
 absence of
 divergence of $\chi^{\cs}_J$  was originally demonstrated by Leggett back in 1965 (Ref.~\onlinecite{Leggett1965}). This
 topic
  has re-surfaced
 recently in the context of
 the discussion about
 a
 $p-$wave Pomeranchuk instability in the spin channel.~\cite{wu:2004,wu:2007,chubukov:2009,Kiselev,Wu2018}

 In the rest of the paper,
  we analyze the interplay between the effects from high- and low-energy fermions
 in some detail. We consider
 a translationally
  and rotationally
  invariant system of fermions with some dispersion $\epsilon_\p$, which is not necessary parabolic (as it would be for a Galilean-invariant system)
 but can be an arbitrary function of $|\k|$.
    We first review the formulation of the
    microscopic theory of a  FL
       in terms of the Ward identities
    associated with conservation laws for
    total charge, spin, and
     momentum. We
    show
     that
     these conservation laws
      give rise to five relations. The first two are
     the original Pitaevskii-Landau relations. They  express $1/Z$ and $m^*/m$ in terms of the vertex function $\Gamma^\omega(p_F,q),$
 in which
 the first
  fermion is on the FS
  while the other is, in general, away from it.
  The other three relations
impose
 the constraints on $\Gamma^\omega(p_F, q)$, one of which directly relates the contributions from low-energy and high-energy fermions
 to each other.
 We then show how these constrains prevent a Fermi surface deformation with the structure of spin current
  and
   charge current order
   order parameters.
      Following that, we review
   a
    diagrammatic derivation of the constraints,
    imposed by
    conservation laws, and
     a diagrammatic calculation
    of the
   charge and spin susceptibilities
    with arbitrary form-factors.
     We argue that,
    for Fermi-surface deformations with
    structures different from
   those
   of charge or spin currents, renormalization
    by high-energy fermions
    reduces the divergence of the corresponding susceptibility at the Pomeranchuk instability
     but does not eliminate it completely,
     i.e., a Pomeranchuk
     instability towards a phase with such order parameter is not forbidden.
     Finally, we present the results of perturbative calculations to second order in
     a four-fermion interaction
     and identify  a
     particular relation involving particle-hole and particle-particle
     polarization bubbles.
     This relation
     allows
     one to re-express the contribution from high-energy fermions as the contribution from the FS,
     and vice versa.

\section{Microscopic theory
  of a
  Fermi liquid}
\label{flt}
\subsection{Pitaevskii-Landau and Kondratenko relations}
Consider a translationally-invariant system of fermions with $H = H_{\text{kin}} + H_{\text{int}}$, where
\beq
H_{\text{kin}} = \sum_{p \alpha}
\epsilon_{\p}
c^\dagger_{\p,\alpha} c^{\phantom{\dagger}}_{\p,\alpha} \label{H}
\eeq
(with chemical potential included into $\epsilon_{\p}$)
 and
 \beq
H_{\text{int}} = \sum_{\k,\p,\q,\alpha,\beta} U(\q) c^\dagger_{\k+\q/2,\alpha} c^\dagger_{\p-\q/2,\beta}c^{\phantom{\dagger}}_{\p+\q/2,\beta} c^{\phantom{\dagger}}_{\k-\q/2,\alpha}.
\notag
\eeq
We
also
 assume that rotational invariance is
intact,
i.e., that the dispersion $\epsilon_{\p}$ depends on
the magnitude of $\p$ but not on its direction,
and
 $U(\q) = U(|\q|)$.
 However, we do not assume a specific form of $\epsilon_{\p}$. It can be parabolic,
as in $^3$He
and near the $\Gamma$-point of the Brillouin zone in cubic materials,
or  linear in $|\p|$, as in Dirac and Weyl materials, \cite{vafek:2014}
 or else
quadratic at the smallest $\p$ and linear at larger $\p$, as in bilayer graphene.\cite{neto:2009}
For all these cases, we assume that
 renormalization of the fermionic properties
by
 interaction
  comes
 predominantly
  from
those momenta which
 are small enough for
 the lattice effects
 to be
 irrelevant.
\footnote{By ``lattice effects'', we mean not only anisotropy but also multiple bands, which are inherent to Dirac and Weyl materials. Our model is applicable to these materials provided that
i) they are doped and ii) the interaction decays on a scale
much smaller than $p_F$, in which case inter-band coupling can be neglected.}
In all
 the
 cases,
 $\epsilon_\p
 \approx p_F (|\p|-p_F)/m$
 near the Fermi momentum,
 where
 $(2m)^{-1}=\left(\pd{\epsilon_{\p}}/\pd{\p^2}\right)\vert_{|\p|=p_F}$.

The propagator of free fermions
 is
\begin{equation}
G_p = \frac{1}{\omega -
\epsilon_\p+i\delta\text{sgn}\epsilon_\p},
\label{new_1_1}
\end{equation}
where
 $p = (\omega, \p)$.
 For interacting fermions
 the
 Landau FL theory states that
\begin{equation}
G_p = \frac{Z}{\omega -
\epsilon_\p(m^*/m)+i\delta\text{sgn}\epsilon_\p} + G_{p,\text{inc}},  \label{new_1}
\end{equation}
where $G_{p,\text{inc}}$ describes incoherent background.
For $\omega \approx
\epsilon_\p(m/m^*)$,  $G_{p,\text{inc}}$ is vanishingly small compared to the first term in (\ref{new_1}).

The microscopic theory of a FL  expresses the quasiparticle residue $Z$ and the effective mass $m^*$ in Eq. (\ref{new_1})  in terms of the bare mass $m$ in (\ref{new_1_1}) and
 a fully renormalized and anti-symmetrized  four-fermion vertex,
$\Gamma^\omega_
{\alpha\beta,\gamma\delta} (p,q)$,
 where $\alpha\dots\delta$ denote the spin projections.
 This vertex
 describes the
interaction
 between fermions with incoming $D+1$-momenta
 $p=(\omega,\p)$ and $q=(\omega',\q)$ and outgoing
 momenta
 $p_1=(\omega_1,\p_1)$ and $q_1=(\omega_1',\q_1)$, taken in the limit of strictly zero
momentum transfer and
vanishingly small
 energy transfer, i.e., for $|\mathbf{p}_1|=|\mathbf{p}|$, $|\mathbf{q}_1|=|\mathbf{q}|$, $\omega_1\to \omega$, and $
\omega_1^{\prime}\to \omega^{\prime}$.
 To first order in $U(\q)$,  $\Gamma^\omega_{\alpha\beta,
\gamma\delta} = U(0) \delta_{\alpha \gamma} \delta_{\beta \delta} - U(|\p-\q|) \delta_{\alpha \delta} \delta_{\beta \gamma}$.
 In general, $\Gamma^\omega_{\alpha\beta,\gamma\delta} (p,q)$ contains contributions only from high-energy fermions and can be computed by setting both energy and momentum transfer to zero.

The relations between $Z$ and $m^*/m$ and the vertex $\Gamma^\omega$ follow from the identities for the
 derivatives of the fermionic Green's functions.
 These identities are
associated with
conservation of
total charge (or, equivalently, total number of fermions),
total spin, and
total momentum.
 The set  is over-complete in the
  sense that the identities associated with
  charge conservation alone
  allow one
  to express
  $Z$ and $m^*/m$
via the vertex function. The
 remaining
  identities
  place
  constraints on the vertex function (see below).

 The identities associated with charge conservation were first derived
 by Pitaevskii and Landau, and it is
 appropriate to
  call
  them
 Pitaevskii-Landau (PL) relations~\cite{Lifshitz1980,agd,Pitaevskii1960}. (For the history of deriving the PL relations, see Ref.~\onlinecite{Pitaevskii2011}.) Although these relations were derived originally for a quadratic dispersion, one can readily obtain them in a form valid for arbitrary $\epsilon_\p$:
\begin{widetext}
\begin{subequations}
\bea
&&\frac{\partial G_p^{-1}}{\partial \omega} = \frac{1}{Z} = 1 - \frac{i}{2}  \sum_{\alpha\beta}
\int \Gamma^\omega_{\alpha\beta,\alpha\beta}  (p_F,q) (G^2_q)^\omega
\frac{d^{D+1} q}{(2\pi)^{D+1}}, \label{2_a} \\
&&{\bf p}_F\cdot\frac{\partial G^{-1}_p}{\partial {\bf p}} = - \frac{p^2_F}{m^*Z}
 = -\frac{p^2_F}{m} + \frac{i}{2}  \sum_{\alpha\beta}
\int \Gamma^k_{\alpha\beta,\alpha\beta}  (p_F,q) \frac{{\bf p}_F\cdot {\bf q}}{m} \frac{\partial \epsilon_{\q}}{\partial \epsilon^{\text{par}}_{\q}}
(G^2_q)^k \frac{d^{D+1} q}{(2\pi)^{D+1}},  \label{2_b}
\eea
\end{subequations}
\end{widetext}
where
$q=(\omega,\q)$,
$\p_F = p_F\hat p$, $\hat p$ is a unit vector
 along $\p$,
 and
$\epsilon^
{\text{par}}
_{\q} =
\q^2/(2m) -E_F
$
with the same $m$  as in (\ref{new_1_1}).

 In Eqs.~(\ref{2_a}) and (\ref{2_b})
 the object  $\left(G^2_q\right)^\omega$ is  the product of two Green's
functions with the same momenta and infinitesimally close frequencies,
 and $\Gamma^k_{\alpha\beta,\alpha\beta}$ is the vertex  in the limit of zero
frequency transfer and vanishing momentum transfer.
In similarity to $\Gamma^\omega$ and $\Gamma^k$, $\left(G^2_q\right)^\omega$ contains only contributions from high-energy fermions and can be
 replaced by just the square of the Green's function, $G^2_q$, while $\left(G^2_q\right)^k$
 contains 
 an 
 additional contribution from fermions at the FS. The vertices $\Gamma^k$ and $\Gamma^\omega$ are
related
 to each other
 by an integral equation
\begin{eqnarray}
&&\Gamma^k_{\alpha \beta,\alpha\beta} (p,q) = \Gamma^\omega_{\alpha
\beta,\alpha\beta} (p,q) \notag \\
&& -\frac{k^ {D-2}_F Z^2 m^*}{(2\pi)^D}\sum_{\xi,\eta} \int
\Gamma^\omega_{\alpha \xi,\alpha\eta} (p,q^{\prime}) \Gamma^k_{\eta
\beta,\xi\beta} (q^{\prime},q) d \Omega_{\q^{\prime}}.\nn\\  \label{n_1}
\end{eqnarray}
where  $d\Omega_\q$ is the infinitesimally small solid angle
around vector $\q$,
  while $\left(G^2_q\right)^k$ is related to $\left(G^2_q\right)^\omega$
by
\begin{equation}
\left(G^2_q\right)^k -\left(G^2_q\right)^\omega \equiv \delta G^2_q= - \frac{
2\pi i Z^2 m^*}{p_F} \delta(\omega) \delta(|\mathbf{q}| -p_F).  \label{n_2}
\end{equation}

Equations (\ref{2_a}) and (\ref{2_b})
 are the most general results for $Z$ and $m^*/m$
 Each equation
contains the integrals over
 the
 intermediate states with momenta $q$ not confined to the FS.
Therefore,
in general,
renormalizations of both $Z$ and $m^*/m$ come from
high
 energies.
  Substituting Eqs.~(\ref{n_1}) and (\ref{n_2}) into Eq.~(\ref{2_b}) we obtain, after
some manipulations,
\begin{equation}
\frac{m^*}{m} = Q  \left(1 - \frac{Z^2 p_F^{(D-2)} m^*}{2 (2\pi)^D} \int
\Gamma^\omega_{\alpha\beta,\alpha\beta} (p_F,q_F) \frac{\mathbf{p}_F\cdot
\mathbf{q}_F}{p^2_F}~ d \Omega_\q\right),
\label{2_bbb}
\end{equation}
where
\begin{equation}
Q = \frac{1 - \frac{i}{2}  \sum_{\alpha\beta}
\int \Gamma^\omega_{\alpha\beta,\alpha\beta}  (p_F,q) (G^2_q)^\omega
\frac{d^{D+1} q}{(2\pi)^{D+1}}}{1 - \frac{i}{2}  \sum_{\alpha\beta}
\int \Gamma^\omega_{\alpha\beta,\alpha\beta}  (p_F,q) (G^2_q)^\omega
\frac{{\bf p}_F\cdot {\bf q}}{p^2_F} \frac{\partial \epsilon_\q}{\partial \epsilon^{\text{par}}_{\q}}
\frac{d^{D+1} q}{(2\pi)^{D+1}}}.
\label{18_1}
\end{equation}
 The integral in the round brackets in (\ref{2_bbb})
goes only over $\Omega_\q$, which implies that this contribution to $m^*/m$ comes solely
from fermions  on the FS.  On the other hand, the factor of $Q$ comes from
high-energy
 fermions.
 Similarly
 to $Q$,
 renormalization of $Z$ in (\ref{2_a}) also comes from high energies.

For
an
$SU(2)$
-invariant FL,  the vertex function can be decoupled into
the
density (charge) and spin components as
\begin{equation}
\Gamma^\omega_{\alpha\beta,\gamma\delta} (p,q) =   \delta_{\alpha\gamma}\delta_{\beta\delta}  \Gamma^c (p,q) +  \boldsymbol{\sigma}_{\alpha\gamma}\cdot \boldsymbol{\sigma}_{\beta\delta} \Gamma^s (p,q),
\label{wed_22}
\end{equation}
 where $\boldsymbol{\sigma}$ is a vector of Pauli matrices.   Because $\sum_{\alpha \beta}\Gamma^\omega_{\alpha \beta,\alpha \beta} (p,q) = 2 \Gamma^c (p,q)$,
   relations (\ref{2_a}) and (\ref{2_b}) contain only
   the
   charge components
  of $\Gamma$:
\begin{widetext}
\begin{subequations}
\bea
&&\frac{\partial G_p^{-1}}{\partial \omega} = \frac{1}{Z} = 1 - 2i
\int \Gamma^c (p_F,q) (G^2_q)^\omega
\frac{d^{D+1} q}{(2\pi)^{D+1}}, \label{2_aa} \\
&&{\bf p}_F\frac{\partial G^{-1}_p}{\partial {\bf p}} = - \frac{p^2_F}{m^*Z}
 = -\frac{p^2_F}{m} + 2i
\int \Gamma^{c}_k (p_F,q) \frac{{\bf p}_F\cdot {\bf q}}{m} \frac{\partial \epsilon_{\q}}{\partial \epsilon^{\text{par}}_{\q}}
(G^2_q)^k \frac{d^{D+1} q}{(2\pi)^{D+1}},  \label{2_bb}
\eea
\end{subequations}
\end{widetext}
 where $\Gamma^{c}_k$ is the charge component of $\Gamma^k$.

Due to rotational invariance, $\Gamma^c$ and $\Gamma^s$  can be
expanded
 in  partial components with different angular momenta
\begin{equation}
\Gamma^{c(s)} (p,q) = \sum_l \Gamma^{c(s)}_l (p,q) K_l,
\end{equation}
 where $K_l$ are the normalized
  angular momentum eigenfunctions,  which depend on the angle $\theta$ between ${\bf p}$ and ${\bf q}$. In 3D, $K_l  = K_l (\theta) = (2l+1) P_l (\theta)$, where $P_l (\theta)$ are Legendre polynomials. In 2D, $K_l (\theta) = \alpha_l \cos{l \theta}$, where $\alpha_0 =1$ and $\alpha_{l>0} =2$.     For the vertex function on the FS,
 the
 partial components $\Gamma^{c(s)}_l$ are
 related to
 the
 Landau parameters $F^{c(s)}_l$,
 introduced in the phenomenological FL theory, via
\begin{equation}
 F^{c(s)}_l = \frac{Z^2 p^{D-2}_F m^*}{\pi^{(D-1)}}\Gamma^{c(s)}_l.
 \label{fr_1}
 \end{equation}
 Substituting this relation into
Eq.~(\ref{2_bbb}), we obtain
 \begin{equation}
 \frac{m^*}{m} = (1 + F^c_1) Q.
 \label{fr_2}
 \end{equation}
In
the
phenomenological
 FL theory, $F^c_1$ is considered as an input, and Eq.~(\ref{fr_2}) relates $m^*/m$ to this parameter.
In
the
 microscopic FL theory, $F^c_1$ is obtained from the vertex function and by itself contains $m^*$ via (\ref{fr_1}). Equation (\ref{fr_2}) then should be viewed as an equation
 for
  $m^*/m$, which one has to solve, if the goal is to express $m^*/m$ in terms of $\Gamma^\omega$.

 Pitaevskii and Landau derived also an additional relation associated with
 momentum conservation
 \bea
 \frac{1}{Z} &=& 1 - \frac{i}{2}
\int
\sum_{\alpha\beta}
\Gamma^\omega_{\alpha\beta,\alpha\beta}  (p_F,q) (G^2_q)^\omega \frac{
{\bf p}_F \cdot{\bf q}}{p^2_F}
\frac{d^{D+1} q}{(2\pi)^{D+1}} \nonumber \\
&&= 1 - 2i
\int \Gamma^c  (p_F,q) (G^2_q)^\omega \frac{
{\bf p}_F \cdot{\bf q}}{p^2_F}
\frac{d^{D+1} q}{(2\pi)^{D+1}}.\nn\\
\label{2_c}
\eea
 It is similar to Eq.~(\ref{2_a}), but  contains
 an
 extra momentum-dependent piece in the r.h.s.  Although
 \eqref{2_c}
 was
 derived
 originally
 for a Galilean-invariant FL,~\cite{Landau1,agd,Lifshitz1980,Pitaevskii1960}
 it holds
 for arbitrary $\epsilon_\p$ (see below).

 Equations (\ref{2_a}) and (\ref{2_c})  show that $1/Z$ can be expressed via $\Gamma$ in two different ways. This obviously places a constraint on the
  vertex function,
  namely, it must satisfy
  \beq
  \int \Gamma^c (p_F,q) (G^2_q)^\omega \left(1- \frac{
{\bf p}_F \cdot{\bf q}}{p^2_F}\right)
\frac{d^{D+1} q}{(2\pi)^{D+1}} =0.
 \label{extra_1}
 \eeq
 Equation (\ref{2_c})
  allows one to
 re-write
 $Q$ in Eq. (\ref{18_1}) in a more transparent way, as
\begin{equation}
Q = \frac{1 - 2i
\int \Gamma^c  (p_F,q) (G^2_q)^\omega
\frac{{\bf p}_F\cdot {\bf q}}{p^2_F} \frac{d^{D+1} q}{(2\pi)^{D+1}}}{1 - 2i
\int \Gamma^c (p_F,q) (G^2_q)^\omega
\frac{{\bf p}_F\cdot {\bf q}}{p^2_F} \frac{\partial \epsilon_\q}{\partial \epsilon^{\text{par}}_{\q}}
\frac{d^{D+1} q}{(2\pi)^{D+1}}}.
\label{18_1_a}
\end{equation}
We immediately see that
 if the dispersion is parabolic within the domain of integration over $q$, i.e.,
$\epsilon_{\p} = \epsilon^{\text{par}}_{\p}$, $Q=1$.
In this case,
 mass renormalization
 comes solely from fermions at the FS~\cite{Landau1,agd,Lifshitz1980}:
\begin{equation}
 \frac{m^*}{m} = 1 + F^c_1.
 \label{fr_2_gi}
 \end{equation}
This result
 was
 originally derived in the phenomenological
FL theory with the help of  Galilean boost.~\cite{Landau1,agd}
Later on, it was shown to be also valid for a Lorentz-invariant relativistic FL.\cite{baym:1976}

A Green's function  at arbitrary $\omega$ and $\p$ can expressed via
the self-energy as $G^{-1} (\omega, \epsilon_{\mathbf{p}}) = \omega -
\epsilon_\p + \Sigma_{\mathrm{FL}} (\omega, \epsilon_{\mathbf{p}})$.
Near the FS, the Green's function must reduce to the first term in Eq.~\eqref{new_1}. This implies that the self-energy must scale linearly with $\omega$ and $\epsilon_\p$ when both  these variables are small.
Combining Eqs. (\ref{2_a}), (\ref{2_bbb}), and (\ref{2_c}),  one
can construct the
self-energy to first order in $\omega$ and $\epsilon_{\mathbf{p}}$.
 After some
manipulations, we obtain

\begin{equation}
\Sigma_{\mathrm{FL}} (\omega, \epsilon_{\mathbf{p}}) = Q_1 \left(\omega -
\epsilon_\p
\right) +  Q_2
\epsilon_\p,
\label{15_a}
\end{equation}
where
\begin{widetext}
\bea
&& Q_1 =
\frac 1Z-1=
  -2i
\int \Gamma^c  (p_F,q)
\left(G^2_q\right)^\omega
\frac{d^{D+1} q}{(2\pi)^{D+1}}, \nonumber \\
&& Q_2 =
\frac 1Z\left(1-\frac{m}{m^*}\right)=
 \frac{1}{Z} \left[1-\frac{1}{Q}
\left(1 - \frac{Z^2 p_F^{(D-2)} m^*}{2 (2\pi)^D} \int
\Gamma^\omega_{\alpha\beta,\alpha\beta} (p_F,q_F) \frac{\mathbf{p}_F\cdot
\mathbf{q}_F}{p^2_F}~ d \Omega_\q\right)^{-1}
\right].
\label{15}
\eea
\end{widetext}

Later on,
Kondratenko~\cite{kondratenko:1964,kondratenko:1965}
derived the relations between $1/Z$, $m^*/m$ and $\Gamma^\omega$ associated with
 conservation of total spin.
The relations are the same as Eqs. (\ref{2_a}) and (\ref{2_b}) but
 contain
extra Pauli matrices, which
 select the spin components of $\Gamma^{\omega}$
 and  $\Gamma^{k}$
 \begin{widetext}
 \begin{subequations}
\bea
&&\frac{\partial G_p^{-1}}{\partial \omega} = \frac{1}{Z} = 1 - 2i
\int \Gamma^s (p_F,q) (G^2_q)^\omega
\frac{d^{D+1} q}{(2\pi)^{D+1}}, \label{2_as} \\
&&{\bf p}_F\frac{\partial G^{-1}_p}{\partial {\bf p}} = - \frac{p^2_F}{m^*Z}
 = -\frac{p^2_F}{m} + 2i
\int \Gamma^s_k
 (p_F,q) \frac{{\bf p}_F\cdot {\bf q}}{m} \frac{\partial \epsilon_{\q}}{\partial \epsilon^{\text{par}}_{\q}}
(G^2_q)^k \frac{d^{D+1} q}{(2\pi)^{D+1}}.  \label{2_bs}
\eea
\end{subequations}
\end{widetext}
where $\Gamma^{s}_k$ is the spin component of $\Gamma^k$.
 Combining Eqs.~(\ref{2_aa}) and (\ref{2_as}),
 and Eqs.~(\ref{2_bb}) and (\ref{2_bs}), we obtain
  two
  additional
  constraints on the vertex function, which relate
the
charge and spin components of $\Gamma^\omega$
 and $\Gamma^k$
to each other via~\cite{Dzyaloshinski1976,Chubukov2014}
 \begin{flalign}
   \int \left(\Gamma^c (p_F, q) -\Gamma^s (p_F, q)\right)  \left( G^{2}_q\right)^\omega
     \frac{d^{D+1} q}{(2\pi)^{D+1}}
     =0, \nonumber \\
 \int \left( \Gamma^{k,c} (p_F,q) - \Gamma^{k,s} (p_F,q)\right)  \frac{{\bf p}_F\cdot {\bf q}}{m} \frac{\partial \epsilon_{\q}}{\partial \epsilon^{\text{par}}_{\q}}
 (G^2_q)^k \frac{d^{D+1} q}{(2\pi)^{D+1}}. \nn \\
 =0
 \label{su_1}
 \end{flalign}
We emphasize that the integrals in (\ref{extra_1}) and (\ref{su_1})
are determined by
high-energy
fermions.  
These equations set the conditions on
the
input parameters
of the phenomenological
FL theory.

\subsection{Pitaevskii-Landau and Kondratenko relations as Ward identities}
\label{sec:lp-relations-from}

The PL relations, Eqs.~\eqref{2_aa}, \eqref{2_bb}, and \eqref{2_c}, and
the
Kondratenko relations, Eqs.~\eqref{2_as} and \eqref{2_bs},
can be recast into
 a more
 compact form by adopting
 the
 general
  formalism of Ward identities, in which
conservation laws are expressed as relations between certain vertex functions and Green's functions.  To obtain these relations, we introduce three momentum and frequency-dependent operators, bilinear in fermions, which we associate with conserved  "charges". In our case these conserved charges  are
charge, spin,
and momentum densities, which are
defined as
\begin{subequations}
   \begin{align}\label{eq:rho-def_n}
  \r^c(\q) &= \sum_{\p,\alpha} c^{\dagger}_{\p-\q/2,\alpha}c^{\phantom{\dagger}}_{\p+\q/2,\alpha}, \\
  \r^s (\q) &= \sum_{\p,\alpha\beta} c^{\dagger}_{\p-\q/2,\alpha}\sigma^z_{\alpha\beta}c_{\p+\q/2,\beta}^{\phantom{\dagger}}, \\
   \boldsymbol{\r}^{{\text{mom}}} (\q) &= \sum_{\p,\alpha} {\bf p} c^{\dagger}_{\p-\q/2,\alpha}c_{\p+\q/2,\alpha}^{\phantom{\dagger}}.
\end{align}
\end{subequations}
Due to spin-rotational invariance, we can consider only one component of the spin density, e.g., along the $z$-axis.
For each conserved quantity
there exists a continuity equation of the form
\bea
  \label{eq:cont-operators}
  \frac{\partial \r^{c,s}(\q)}{\partial t} &=& -i \q \cdot \boldsymbol{\r}\cs_{J} (\q),\nn\\
   \frac{\partial\r_l^{\text{mom}}(\q)}{\partial t}&=&-i\sum_kq_k\r^{\text{mom}}_{J,lk}(\q),
\eea
where
 $\boldsymbol{\r}_{J_{c,s}}$ are the operators of charge and spin currents,
and
$\r^{\text{mom}}_{J,lk}(\q)$ is the momentum current , i.e., the energy-momentum tensor.
Both relations in Eq.~\eqref{eq:cont-operators} are exact for a quadratic dispersion, i.e., for
 a
 Galilean-invariant system, and valid to lowest order in $|\q|/p_F$ otherwise.

Using Heisenberg equations of motion for the operators of charge and spin densities,
one can verify
that the corresponding currents are also bilinear in fermions:
\begin{align}\label{eq:rho-def_n1}
 \boldsymbol{\r}^c_{J}  (\q) &= \sum_{\p,\alpha} \frac{\partial \epsilon_\p}{\partial \p}c^{\dagger}_{\p-\q/2,\alpha}
  c^{\phantom{\dagger}}_{\p+\q/2,\alpha},\nn \\
 \boldsymbol{\r}^s_{J} (\q) &= \sum_{\p,\alpha\beta} \frac{\partial \epsilon_\p}{\partial \p}c^{\dagger}_{\p-\q/2,\alpha}
 \sigma^z_{\alpha\beta}
  c_{\p+\q/2,\beta}^{\phantom{\dagger}}.
\end{align}
The situation with the
the energy-momentum tensor is more subtle.
In the phenomenological FL theory, this tensor has the usual hydrodynamic form
$\rho^{\text{mom}}_{J,ij}=\int p_i(\pd\epsilon_\p/\partial p_j) n_\p d^Dp/(2\pi)^D$, where $n_\p$ is the quasiparticle occupation number.\cite{baym:book} This might suggest that
the second-quantized form of the energy-momentum
 tensor
  is also a bilinear, similar to those in Eq.~\eqref{eq:rho-def_n1}, but with $\pd\epsilon_\p/\partial\p$ replaced by $p_i(\pd\epsilon_\p/\partial p_j)$.
An explicit calculation indeed shows that the commutator $[H_{\text{kin}},\boldsymbol\r^{\text{mom}}(\q)]$ does yield a bilinear part of the energy-momentum tensor.
However,
at $\q\neq 0$ there is also another, quartic in fermions part which comes from $[H_{\text{int}},\boldsymbol\r^{\text{mom}}(\q)]$
(Refs.~\onlinecite{Wu2018,Woelfle_private}).

Because of the latter part,
the energy-momentum tensor cannot be expressed
via
a purely
bilinear combination of fermions. In what follows, we will use the continuity
equation for the momentum density only at $\q=0$, when the complications due
to the quartic part of the energy-momentum tensor do not arise.

\begin{figure}
  \includegraphics[width=\hsize,clip,trim=0 40 0 40]{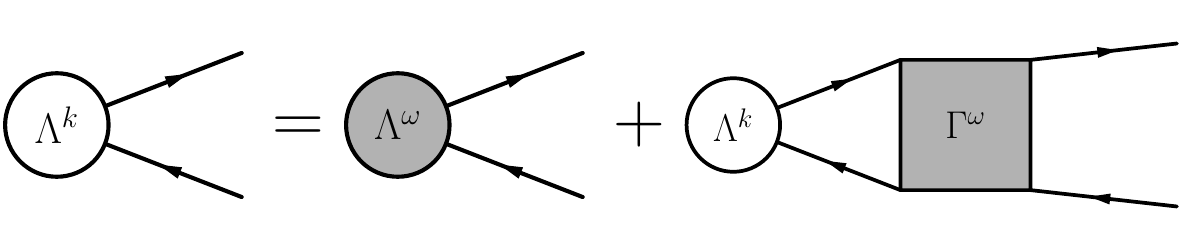}
  \caption{
 Graphical representation of the relation between three-leg vertices $\Lambda^k$ and $\Lambda^\omega$.
  }\label{fig:3-4vertex}
\end{figure}

Graphically,
we can
associate
each conserved
charge
and its corresponding current
with fully renormalized
three-leg vertices
  $\Lambda\cs (p, q)$, $\boldsymbol{\Lambda}^{{\text{mom}}}$, $\boldsymbol{\Lambda}\cs_{J}(p,q)$,
  and $\Lambda^{\text{mom}}_{J,ij}$, as shown in Fig. \ref{fig:3-4vertex}.  We define $\boldsymbol{\Lambda}\cs_{J} (p,q)$ and $\boldsymbol{\Lambda}^{{\text{mom}}}$ without overall form-factors of $\partial \epsilon_\p/\partial \p$  and ${\bf p}$, respectively.  With this definition, all vertices
  are equal to unity
   for free fermions.

To derive the Ward identities connecting the three-leg vertices for charges and currents, we follow  Engelsberg and Schrieffer \cite{engelsberg:1963,Fabrizio2013}
and compute the time derivative of a time-ordered combination
\begin{equation}
  \label{eq:vertex-time-rep}
  \langle T_t c(\p+\frac \q 2,t_1)c^\dagger(\p-\frac \q 2,t_2) \r^n (\q,t)\rangle,
\end{equation}
where $n=c(s)$.
 It is easy to see that the derivative $\pd/\pd_t$  yields a term proportional to $\pd \hat\rho^n /\pd_t = -i\q\cdot \hat{\boldsymbol\rho}^n_{J}$ and additional terms proportional to $\delta(t-t_1)$
and
$\delta(t-t_2)$,
 which originate from
differentiating
 the time-ordering operator $T_t$.  Using standard manipulations and
 Fourier transforming in time,
 one finds \cite{engelsberg:1963,Fabrizio2013}
 \begin{equation}
  \label{eq:ward-identity}
  \omega
  \Lambda^n
  (p,q)  - \q \cdot
  \frac{\partial\epsilon_\p}{\partial\p}
  \Lambda^n_{J}
  (p,q) =
  G^{-1}_
  {p+ q/2}
   - G^{-1}_
   {p - q/2},
\end{equation}
where
$\Lambda^n_{J}\equiv \hat p\cdot \boldsymbol{\Lambda}^n_{J}$
 and $q=(\omega,\q)$.

Similarly to $\Gamma^\omega$ and $\Gamma^k$ in Eqs.~\eqref{2_a} and \eqref{2_b},
it is  convenient to define the
vertices  $\Lambda^{\omega} (p)
 = \lim_{\omega\to 0,|\q|\to 0,|\q|/\omega\to 0}\Lambda
 (p,q)$ and $\Lambda^{k}
 (p) =\lim_{\omega\to 0,|\q|\to 0,\omega/|\q| \to 0} \Lambda
 (p,q)
 $.
In
analogy with
$\Gamma^k$
and $\Gamma^\omega$,
renormalization of
$\Lambda^\omega$  comes from high-energy fermions
 while
 renormalization
of $\Lambda^k$
comes from both high- and low-energy fermions.
 The relation between the vertices $\Lambda^k_{n,J_n}$ and $\Lambda^\omega_{n,J_n}$ follows from Fig.~\ref{fig:3-4vertex},
 and is  similar to the one which relates $\Gamma^k$ and $\Gamma^\omega$ in Eqs. (\ref{n_1}) and (\ref{n_2}).
 For example, the charge and spin vertices satisfy
\bwt
\begin{align}
  \Lambda^{k,n}(p
  )\sigma^{
  n}_{\beta\beta} &= \Lambda
  ^{\omega,n}(p
  )\sigma^{
  n}_{\beta\beta}
  -
  \frac{k^{D-2}_F Z^2 m^*}{(2\pi)^D}\sum_{\xi,\eta} \int
  \Lambda^{k,n} (p^\prime_F
  )\sigma^{
  n}_{\xi\eta} \Gamma^\omega_{\eta
  \beta,\xi\beta} (p^{\prime}_F,p) d \Omega_{\p^{\prime}_F},  \label{n_1a}
\end{align}
\ewt
 with $n=(c,s)$
 and similarly for the vertices of corresponding currents.
Here, $\sigma^{c}$ denotes the identity matrix and
$\sigma^{s}=\sigma^z
$. Projecting
$p$
onto the
FS, using (\ref{n_1a}) to express $\Lambda^k$ in terms of $\Lambda^\omega$ and $\Gamma^\omega$,  and taking separately the limits $
\omega =0$ and $\q = 0$,
we obtain the following four identities
\begin{subequations}
\begin{align}
  \Lambda^c
  Z &= 1,\quad\quad
      \Lambda^s
      Z = 1; \label{eq:ward-LZ-cs} \\
  \frac{m^*}{m}\Lambda^c_{J} Z  &= 1 + F_1^{c},\quad\quad \frac{m^*}{m}\Lambda^s_{J}Z = 1 + F_1^{s}, \label{eq:ward-mLZ-cs}
\end{align}
\end{subequations}
where
we re-defined
$\Lambda^{n}\equiv \Lambda^{\omega,n}$ and $\Lambda^{n}_J\equiv \Lambda^{\omega,n}_J$
for brevity.

As we said,  the
energy-momentum tensor
is not expressed as a bilinear combination of fermions, hence
it cannot be expressed graphically as in Fig.~\ref{fig:3-4vertex}.
Nevertheless, we can
still use Eq.~(\ref{eq:ward-identity}) for $\boldsymbol{\r}_{{\text{mom}}}$ at $\q=0$, when
the momentum-energy tensor
does not contribute.
Taking this limit, we obtain the fifth identity
\beq
\Lambda^{{\text{mom}}} Z = 1, \label{eq:ward-LZ-mom}
\eeq
where $\Lambda^{{\text{mom}}}=\hat p\cdot \boldsymbol{\Lambda}^\omega$.

We now go back to the PL relations. We easily identify the first two PL relations
 \eqref{2_a} and \eqref{2_b},
as
Eqs. \eqref{eq:ward-LZ-cs} and \eqref{eq:ward-mLZ-cs} for $\Lambda_c$ and $\Lambda^c_{J}$. Indeed, the r.h.s. of Eqs.
\eqref{2_a} and
\eqref{2_b} are just  the expressions for the
three-leg
 vertices $\Lambda_c$ and $\Lambda^c_{J}$, as can be seen directly from Fig.~\ref{fig:3-4vertex}.
Explicitly,
\bea
&& \Lambda^c = 1 - 2i
\int \Gamma^c  (p_F,q) (G^2_q)^\omega \frac{d^{D+1} q}{(2\pi)^{D+1}}, \label{k_c} \\
&& \Lambda^c_{J}  = 1 - 2i
\int \Gamma^c (p_F,q) (G^2_q)^\omega
\frac{{\bf p}_F\cdot {\bf q}}{p^2_F} \frac{\partial \epsilon_\q}{\partial \epsilon^{\text{par}}_{\q}}
\frac{d^{D+1} q}{(2\pi)^{D+1}}.  \nonumber
\eea
The factor of $Q$ in Eq. \eqref{18_1}, which incorporates high-energy contributions to $m^*/m$, is then equal to $(\Lambda^c_{J} Z)^{-1}$. Similar considerations apply to the
 Kondratenko
 relations in the
spin channel, Eqs. \eqref{2_as} and \eqref{2_bs}.
In this case,
we have
\bea
&& \Lambda^s = 1 - 2i
\int \Gamma^c  (p_F,q) (G^2_q)^\omega \frac{d^{D+1} q}{(2\pi)^{D+1}}, \label{k_s} \\
&&\Lambda^s_{J}  = 1 - 2i
\int \Gamma^s (p_F,q) (G^2_q)^\omega
\frac{{\bf p}_F\cdot {\bf q}}{p^2_F} \frac{\partial \epsilon_\q}{\partial \epsilon^{\text{par}}_{\q}}
\frac{d^{D+1} q}{(2\pi)^{D+1}}. \nonumber
\eea

The additional PL relation, Eq.~\eqref{2_c}, is identical to  Eq.~\eqref{eq:ward-LZ-mom}, i.e., the r.h.s. of \eqref{2_c} is
 just the definition of $\Lambda^{{\text{mom}}}$:
\beq
 \Lambda^{{\text{mom}}} = 1 - 2i
\int \Gamma^c  (p_F,q) (G^2_q)^\omega \frac{
{\bf p}_F \cdot{\bf q}}{p^2_F}
\frac{d^{D+1} q}{(2\pi)^{D+1}}.
\label{k_mom}
\eeq
Note that Eq.~\eqref{2_c}
is valid both for
Galilean-
 and non-Galilean-invariant systems,
 as long as
the total momentum is a conserved.

We re-iterate that Eqs.~\eqref{eq:ward-LZ-cs} and \eqref{eq:ward-LZ-mom} express $Z$ in three different ways, and
thus
 place constraints on the vertex functions, Eqs.~\eqref{k_c} and \eqref{k_mom}. Equation \eqref{eq:ward-mLZ-cs} relates the product of
 $\Lambda\cs_J$, $Z$, and $m^*/m$
to $1+F_1\cs$.
For a non-Galilean-invariant system,
all the quantities on the l.h.s.
come, at least
partially, from
high-energy
fermions,
while $F_1\cs$
is proportional to
the $l=1$ component of the interaction vertex between low-energy
 fermions.
 
\subsection{Implication of conservation laws for Pomeranchuk instabilities in a Fermi liquid}
\label{sec:disc-impl-cons}

   A Pomeranchuk instability is a spontaneous development of a long-range order in
the
spin or charge channel,
which occurs
when
the
fermion-fermion interaction reaches a critical value.
A distinctive feature of a Pomeranchuk instability is that it breaks either rotational symmetry of the FS or its topology leaving translational
symmetry intact. For example, a ferromagnetic (Stoner) transition is a Pomeranchuk instability, while an antiferromagnetic transition is not.

The order parameters associated with Pomeranchuk instabilities are bilinear in fermions.
Examples of such order parameters were already presented in Eqs. (\ref{eq:rho-def_n}) and (\ref{eq:rho-def_n1}).
 More general
 order
parameters
with angular momentum $l$
in the charge and spin channels can be defined as
\begin{align}\label{eq:rho-def}
  \r^{c}_l (\q) &= \sum_{\p,\alpha} \lambda^c_l (\p) c^{\dagger}_{\p-\q/2,\alpha}c_{\p+\q/2,\alpha}, \\
  \r^{s}_l (\q) &= \sum_{\p,\alpha\beta}  \lambda^s_l (\p)  c^{\dagger}_{\p-\q/2,\alpha}
  \sigma^z_{\alpha\beta}c_{\p+\q/2,\beta}.
\end{align}
where $\lambda^{\cs}_l (\p) $ is  a form-factor,
 which
 transforms under rotations according to
its angular momentum channel (i.e., as $1$ for $l=0$, as $\p$ for $l=1$, etc.).
In 2D
\begin{equation}
 \label{eq:form-factor-def}
 \lambda^{c(s)}_l (\p)= \cos(l\phi_{\p}) |\p|^l\times f^{c(s)}_l(|\p|),
\end{equation}
or equivalently with $\sin$ instead or $\cos$.
In (\ref{eq:form-factor-def}),  $f_l(|\p|)$ can be any function.

A Pomeranchuk instability is usually expressed as a condition on
 the
  Landau parameter $F^{\cs}_l$, defined in Eq.~\eqref{fr_1}. Pomeranchuk's original argument \cite{Pomeranchuk} was
that the prefactor of the term in the ground state energy, quadratic in the
 variation of the shape of a FS
  with given  $l, c(s)$,
scales as $1 + F^{c(s)}_l$ and
vanishes when
 $F^{c(s)}_l
 \to
  -1$.
The corresponding susceptibility $\chi^{c(s)}_{l}$ then scales as $1/(1 + F^{c(s)}_l)$ and diverges at $F^{c(s)}_l = -1$.

The susceptibility  $\chi^{c(s)}_{l}$ computed within the random phase approximation (RPA)
 shows just
this behavior,
 i.e.,
\begin{equation}
  \label{eq:susc-RPA}
  \chi^{c(s)}_{l, \text{RPA}} =  \chi_{l,0} \frac{1}{1 + F^{c(s)}_l},
\end{equation}
where $\chi_{l,0}$ is the susceptibility of free fermions,  given by
\begin{align}
  \chi^{c(s)}_{l=0,0}  &= \frac{m}{\pi}  \left(f^{c(s)}_{l=0} (p_F)\right)^2,  \nonumber \\
  \chi^{c(s)}_{l >0} &=  \frac{m}{2\pi}\left(p_F^lf^{c(s)}_{l} (p_F)\right)^2.
        \label{n_14_0}
\end{align}
 The
 integral
  in the
 free
 particle-hole bubble
  is confined to an infinitesimal region around the FS (see the derivation around Eq. \eqref{n_4} below).
Since
 the interaction between fermions on the FS is
parameterized $F_l^{\cs}$,
the appearance of the denominator $1 + F^{c(s)}_l$ in (\ref{eq:susc-RPA}) is the result of
summing up the geometric series of particle-hole bubbles.

Note that
the free-fermion $\chi_{l,0}$ is finite in the static limit $\Omega =0, \q\to 0$
(which is the case in Eq. (\ref{n_14_0})),  but
vanishes,
for
 any $l$,
in the opposite limit
 of
 $\q =0, \Omega \to 0$,
 given that the system is $SU(2)$-symmetric.
  This vanishing is the consequence of the fact that for free fermions any particle-hole order parameter, bilinear in fermions, is a conserved quantity.  At  small but finite $v_F q/\Omega$,  $\chi^{c(s)}_{l,0} (\q, \Omega)$ scales as $(v_F q/\Omega)^2$.

The RPA expression,
 however,
  is not the full result
 for $\chi^{c(s)}_{l}$.
  An exact formula for the  static susceptibility
   was
    obtained by Legget.\cite{Leggett1965} It reads:
\begin{equation}
  \chi_{l}^{c(s)}= \left(\Lambda^{c(s)}_l Z\right)^2  \frac{m^*}{m}\chi^{c(s)}_{l,\text{RPA}}+\chi^{c(s)}_{l,\text{inc}}.
  \label{eq:full}
\end{equation}
Here $\Lambda_l^{\cs}$ is the same
three-leg
 vertex as before but now for arbitrary order parameter with angular momentum $l$. The
first term in \eqref{eq:full}
 is often called the ``quasiparticle contribution''
  because it is finite in the static limit $\Omega =0, \q \to 0$, but vanishes at $\q =0, \Omega \to 0$
  regardless of wether the corresponding order parameter is conserved or not

   Still, we recall that $\Lambda^{c(s)}_l, Z$ and $m^*/m$
   (for a non-parabolic spectrum)
   in the first term
   are the three input
   parameters
   which come, at least
   partially,
    from high-energy fermions.
The
 second
 term,  $\chi_{l,\text{inc}}^{\cs}$,
 is  the contribution
  only
  from
   high-energy fermions.
 In a generic case, this term is not described at all within
 the
 FL theory, and its value does not depend on the order of limits
 $\Omega\to 0$
and $|\q|\to 0.$

We will review
a
diagrammatic derivation of Eq.~\eqref{eq:full} in the next Section. Here, we  focus on the implications of
 Eq.~(\ref{eq:full}) for
 Pomeranchuk instabilities of a FL.

First, Eq. (\ref{eq:full}) shows that there is more than one scenario for the divergence of
the
  susceptibility
 in a given channel.
 In addition to
 the Pomeranchuk scenario
(the vanishing of $1 + F^{\cs}_{l}$),
 the
 susceptibility $\chi_{l}^{c(s)}$ can
also
 diverge if contributions from high-energy fermions give rise to
a divergence of
 $Z\Lambda^{c(s)}_l$ or $\chi^{c(s)}_{l,\text{inc}}$.
 Finally,
$m^*/m$
 for a non-parabolic spectrum may also diverge due to singular contributions from high-energy
 fermions.
 These three scenarios are outside the FL theory.\footnote{
 One example
 of such a divergence in a non-$SU(2)$-symmetric system is the ferromagnetic instability of a FL with Rashba
 spin-orbit coupling, in which case the entire spin susceptibility comes from high-energy
 fermions.\cite{zak:2010,zak:2012,ashrafi:2013}}

Second,  Eq.~(\ref{eq:full}) shows that
 the divergence of $\chi^{c(s)}_{l, \text{RPA}}$ may, in principle, be canceled by the vanishing of its prefactor, $\Lambda_l^{c(s)}Z(m^*/m)$.
If this happens,
  the corresponding susceptibility remains finite
  at $F^{\cs}_{l} =-1$.
  It will be shown below
  that this is the case for
  order parameters which coincide with
  the momentum density,
  and
  charge and spin
  currents.~\cite{Kiselev,Wu2018}

 To see this, we now systematically analyze the implications of the conservation laws for
 the relation between the two terms in
 \eqref{eq:full}. We
 first consider the susceptibilities
 of
  three conserved order parameters - total charge (particle number), total spin, and total momentum.
 (Here and thereafter, a susceptibility of any vector quantity will be understood as a longitudinal
  part of the corresponding tensor.)
  For the first two
  order parameters
  $l=0$ and $\lambda^{\cs}_{l=0} (\p)  =1$,
  while
  for the third one $l=1$ and $\lambda^{\cs}_{l=1}  (\p) =
  p_x$,
  where we choose the $x$-axis to be along $\q$.
   In all three cases, $\Lambda Z=1$ and
$\chi_{\text{inc}} =0$. Consequently, the susceptibilities
of the three conserved quantities
coincide with the RPA expressions, modulo
a factor
of
 $m^*/m$
\begin{gather}
  \chi^c = \frac{m^*}{\pi} \frac{1}{1 + F^c_0},\qquad \chi^s = \frac{m^*}{\pi} \frac{1}{1 + F^s_0},\nn\\
  \chi_{\text{mom}} = \frac{m^* k^2_F}{2\pi} \frac{1}{1 + F^c_1}
  \label{eq:chi-c-s-mom}
\end{gather}
(for definiteness, we  use the explicit forms of $\chi_{l,0}$ in 2D).
 The $l=0$ instability in the charge channel corresponds to phase separation, the one in the spin channel corresponds to ferromagnetism, and the one at $l=1$ signals the emergence of a charge nematic
  order. In a Galilean-invariant system, $m^*/m = 1 + F^c_1$, and an $l=1$ Pomeranchuk instability
 in the charge channel
  does not occur.

Next, we consider the susceptibilities
of charge
and spin
currents,
 i.e., for order parameters with $l=1$ and form factor
$\lambda^{\cs}_{l=1} (\p) = \partial \epsilon_{\p}/\partial
p_x$. We label the corresponding susceptibilities as $\chi^{c(s)}_J$.
Using (\ref{eq:ward-mLZ-cs}), we can re-express Eq. (\ref{eq:full}) for   $\chi^{c(s)}_J$ as
\beq
\chi^{c(s)}_J = \frac{m p^2_F}{2\pi} \frac{m}{m^*} (1 + F^{\cs}_1) +\chi\cs_{J,\text{inc}}.
\label{k_6}
\eeq
We see that the quasiparticle part of the susceptibility of either charge-current or spin-current order parameter actually vanishes
when the corresponding Landau parameter reaches $-1$, i.e.,
 a  Pomeranchuk instability does not show up if one probes it by analyzing
 particular
 susceptibilities
 as specified above.\cite{Kiselev}

Equation (\ref{k_6}) can be also derived explicitly, by expressing the susceptibilities $\chi^{c(s)}$
via
DM
the
time-ordered correlators  of $\r$ at times $t$ and $t'$,
differentiating over $t$ and $t'$, and
 using the continuity equation, \eqref{eq:cont-operators}. This yields~\cite{Leggett1965,Kiselev}
\begin{equation}
  \label{eq:cont-susc-par}
\omega/|\q|)^2\chi^{c(s)}(\q,
\omega) =
\chi^{c(s)}_J(\q,
\omega) - \chi^{c(s)}_J(\q,0).
\end{equation}
 Now we take the limit $\omega \gg  v^*_F q$, keeping both $\omega$ and $\q$ infinitesimally small.
 The incoherent
 parts of $\chi^{c(s)}_J(\q,\omega)$ and  $\chi^{c(s)}_J(\q,0)$
 cancel each other,
  while the quasiparticle
 part of
$\chi^{c(s)}_J(\q,\omega)$ vanishes.  As
a
  consequence, the r.h.s. of (\ref{eq:cont-susc-par}) reduces to
\beq
-\chi^{c(s)}_J(\q,0)=-\frac{m^* v^2_F }{2\pi} \left((Z \Lambda\cs_{J})^2 \frac{1}{1 +  F^{c(s)}_1}\right).
\eeq
The l.h.s. of  (\ref{eq:cont-susc-par}) tends to a constant at  $\omega \gg  v^*_F q$ because $\chi^{c(s)} (\q, \omega) = \frac{m^*}{m}\chi^{c(s)}_{\text{RPA}} (\q,\omega)$ scales as $(|\q|/\omega)^2$ and cancels out the factor $(\omega/|\q|)^2$.  The expression for
 $\chi^{c(s)}_{\text{RPA}} (\q,\omega)$ in this limit,
 obtained
 by Leggett
 in
 Ref.~\onlinecite{Leggett1968},  reads
 \bea
\chi^{c(s)}_{\text{RPA}} (\q,\omega) =
  -\frac{m}{2\pi} \left(\frac{v_F |\q|}{\omega}\right)^2 \left(\frac{m}{m^*}\right)^2 \left(1 + F^{\cs}_{1}\right).\nn\\
 \eea
 Hence, at $\omega \gg  v^*_F q$,
 \beq
 (\omega/|\q|)^2\chi^{c(s)}(\q,\omega) = - \frac{m}{2\pi} \left(\frac{v_F |\q|}{\omega}\right)^2  \left(\frac{m}{m^*}\right) \left(1 + F^{\cs}_{1}\right).
 \eeq
 Substituting this into (\ref{eq:cont-susc-par}) we reproduce Eq.~(\ref{eq:ward-mLZ-cs}):
 \beq
  Z \Lambda\cs_{J} = \left(\frac{m}{m^*}\right) \left(1 + F^{\cs}_{1}\right).
 \label{k_4}
  \eeq
Substituting
 this
 further into (\ref{eq:full}) we reproduce (\ref{k_6}).
 Note that Eq. (\ref{k_4})
 can be
 re-written
 as
 \begin{equation}
  \label{eq:l-l-ratio}
  \frac{\Lambda^c_{J}}{\Lambda^s_{J}} = \frac{1+F_1^c}{1+F_1^s}.
\end{equation}
We see that  the  vanishing of  $1+F_1^{\cs}$ is always associated with
the
 vanishing of the corresponding $\Lambda\cs_{J}$,
 excluding an unlikely case when $1+F_1^{c}$ and $1+F_1^{s}$ vanish simultaneously.

Leggett showed~\cite{Leggett1965}
that there exists another, even stronger constraint on
the
 susceptibilities of
 charge and spin currents.
Namely, the longitudinal sum rule
implies that $\chi^{c}_J$ and $\chi^{s}_J$
are not renormalized by the interaction,
 i.e.,
\begin{equation}
  \label{eq:f-sum-rule}
  \chi^{c}_J = \chi^{s}_J = \frac{m p^2_F}{2\pi}.
\end{equation}
The longitudinal sum rule is analogous to the longitudinal f-sum rule for the imaginary part of the inverse dielectric function \cite{Mahan} and is the consequence of the gauge-invariance of the electromagnetic field. \cite{Ehrenreich1967}

Constraint (\ref{eq:f-sum-rule})
relates $\chi_{J,\text{inc}}$ in \eqref{k_6}
to
the
 Landau parameter $F^{c(s)}_{1}$:
\begin{equation}
\chi\cs_{J,\text{inc}} = \frac{m p^2_F}{2\pi} \left(1 - \frac{m}{m^*}(1+F_1\cs)\right).
\label{mo_1}
\end{equation}
This is yet another
 condition
 on the contribution coming from high-energy fermions.

In a Galilean-invariant FL, $m^*/m = 1 + F^c_{1}$, and Eqs. \eqref{k_6} - \eqref{eq:f-sum-rule} reduce to
\begin{align}
&Z \Lambda^c_{J} =1, & &\chi^c_{J,\text{inc}} = 0 \label{mo_2} \\
&Z \Lambda^s_{J} = \frac{1 + F^s_{1}}{1 + F^c_{1}}, & &\chi^s_{J,\text{inc}} = \frac{m p^2_F}{2\pi} \frac{F_{1}^c-F^s_{1}}{1+ F^c_{
1}}. \label{mo_3}
\end{align}

The fact that $\chi^{c(s)}_{J}$ is constrained by Eqs.~\eqref{k_6} and \eqref{eq:f-sum-rule}  does not imply that a Pomeranchuk transition in the $l=1$ channel can never occur.
 Indeed, these constraints
 do not preclude the system from developing an instability towards a phase
 described by
 an order parameter with a form-factor, which has the same symmetry as the
 charge- or spin-current
  order parameter
 but depends differently on $|\p|$.
 In Eq.~\eqref{eq:form-factor-def} we defined an infinite family of order parameters with a given angular
 momentum $l$,
   specified
   by an overall scalar function $f(|\p|)$.
  The current susceptibilities
  correspond to the choice (see \eqref{eq:rho-def_n1})
\begin{equation}
  \label{eq:lambda-J-cs}
  f^{\cs}_l
  (|\p|) = \frac{1}{m}\frac{\pd\epsilon_{\p}}{\pd \epsilon_{\p}^{\text{par}}}.
\end{equation}
For
order parameters with $f_{l=1}$  different from the equation above,
there are no
  general
 reasons to expect  $\Lambda_{l=1}^{\cs}Z$ to be proportional to
 $1+F_1^{\cs}$ , hence
  a Pomeranchuk instability is expected to
 occur at  $F_1^{\cs} =-1$ (Ref.~\onlinecite{Wu2018}).

  Interestingly enough,
  we can interpret this instability in two different ways, depending on how we
  write the susceptibility $\chi^{c(s)}_{l=1}$.
If we use the original Eq.~(\ref{eq:full})
\begin{align}
  \label{eq:chi-gen-1}
  \chi_{l=1}^{c(s)}& = \frac{m p_f^2}{2\pi}\frac{m^*}{m}Z^2\frac{(\Lambda^{c(s)}_{l=1})^2}{1+F_1^{\cs}} +\chi^{c(s)}_{l=1,\text{inc}}, \\
 \end{align}
  we
  would
  conclude that the instability is determined by the condition on the interaction between fermions on the FS:  $F_1^{\cs} =-1$.
  However,
  re-writing
   $ \chi_{l=1}^{c(s)}$ as
  \begin{align}
  \label{eq:chi-gen-2}
  \chi_{l=1}^{c(s)} &= \frac{m p_f^2}{2\pi} Z \frac{(\Lambda^{c(s)}_{l=1})^2}{\Lambda\cs_{J}} +\chi^{c(s)}_{l=1,\text{inc}} \\
 \end{align}
we
would
conclude that the Pomeranchuk instability is driven by the vanishing of $\Lambda\cs_{J}$, which is determined by
 high-energy fermions.
 This dual interpretation is yet another consequence of the fact that
 conservation of charge and spin imposes the relations between the properties
of  low- and high-energy fermions.

\section{Diagrammatic derivation of Leggett's
 result
for the static susceptibility}
\label{sec:diagr-deriv-legg}

In this section we review a diagrammatic derivation of Eq.~\eqref{eq:full}, closely following the presentation in Ref.~\onlinecite{Wu2018}. The purpose of this derivation is to show that $\Lambda_l^{\cs}$ and $\chi^{\cs}_{l,\text{inc}}$ arise from
high-energy
contributions.
 For definiteness, we consider the
2D case.
 To simplify notations, here and in the next section we
 suppress
 the superscript
 $\omega$ in $\left(G^2_q\right)^\omega$, i.e.,  replace
 $\left(G^2_q\right)^\omega$
 just by $G^2_q$.

\begin{figure}
  \centering
  \includegraphics[width=0.5\hsize,clip, trim=65 70 58 70]{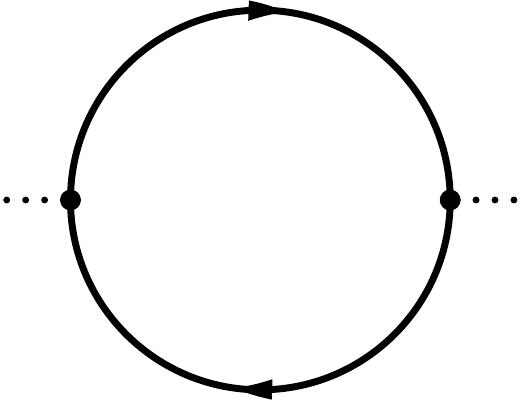}
  \caption{The free fermion susceptibility}
  \label{fig:fig0}
\end{figure}

Let's start with the
free-fermion
susceptibility
 for an
 order parameter with form-factor $\lambda_l^{\cs}$, as in Eq. \eqref{eq:form-factor-def}. The diagrammatic representation of the free-fermion susceptibility $\chi^{c(s)}_{l,0} (q)$ is a bubble
 composed
  of two fermionic propagators (Fig. \ref{fig:fig0}) with
form-factors
$\lambda^{c(s)}_l$
at
  the vertices:
 \begin{equation}
 \chi^{c(s)}_{l,0} (q)=-2
  i\int\frac{d^3
 p}{(2\pi)^3}  \left(\lambda^{c(s)}_l (\bm{p}) \right)^2  G
 _{p + \frac{
 q}{2}} G
 _{p-\frac{q}{2}}.
 \label{n_3_1}
\end{equation}
Here, $G_k$ stands for a free-fermion Green's function, given by Eq.~\eqref{new_1_1},
and  the factor of $2$ comes from spin summation.
  The frequency
  integral in (\ref{n_3_1}) is non-zero only if $\epsilon_{\p+\q/2}$ and $\epsilon_{\p-\q/2}$ have opposite signs
  which, for $|\q|\ll p_F$,
implies that
the integral
over $|\p|$
 comes from
 a narrow region
 near the FS.
 At $T=0$, we have
   \begin{eqnarray}
 &&\chi_{l,0}^{\cs} (q)= - \frac{m}{\pi}
     \left(p_F^lf^{c(s)}_l (p_F)\right)^2
   \nonumber \\
  && \times\int \frac{d\phi_{\p}}{2\pi} (\cos{l \phi_\p})^2 \frac{ v_F |\q| \cos\phi_\p}{\omega - v_F |\q| \cos\phi_\p + i \delta
  \text{sgn}\omega}.
\label{n_4}
\end{eqnarray}
 In  the static limit we
 reproduce Eq.~\eqref{n_14_0}.

The $1/(1 + F^{c(s)}_l)$ dependence of  $\chi^{c(s)}_{l}$ can be reproduced diagrammatically
within RPA. Because the momentum/frequency integration within each bubble is confined to the FS, the dimensionless interaction between the bubbles is exactly $F^{c(s)}_l$.
Re-summing the geometric series,
we reproduce Eq.~\eqref{eq:susc-RPA}.

\begin{figure*}
  \centering
  \includegraphics[width=\hsize,clip,trim= 0 0 0 0]{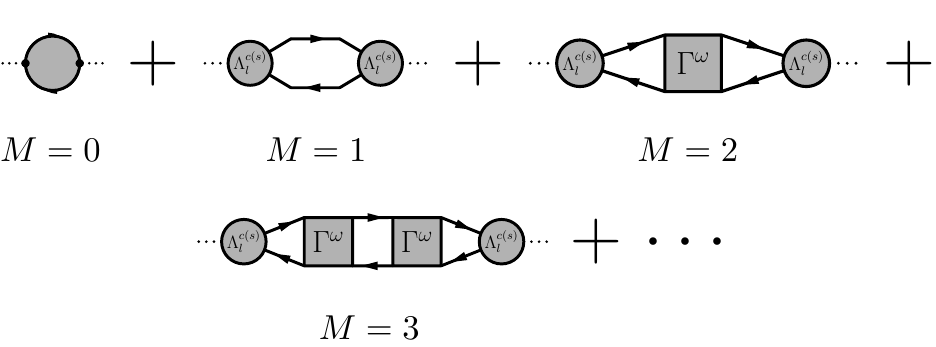}
  \caption{The ladder series of diagrams for the static susceptibility $\chi_l\cs$. The exact $\chi_l$ is represented as a series  of $M=0,1,2,\ldots$
   bubbles comprised of Green's functions with poles
  in
  the
  opposite
  half-planes
  of complex frequency,
  whose contributions are computed close to the FS.
   Gray shading denotes contributions
   from
   high energy fermions, for which the poles in the Green's functions are
  in the
  same half-planes of complex frequency.
  These include the $M=0$ bubble
   (on the far left),
   as well as the vertices $\Lambda_l^{\cs}$
   and $\Gamma^\omega$.
    }
  \label{fig:bubble-series}
\end{figure*}

To obtain
an
 exact expression, we need to go beyond RPA.
To this end,
 we note that a diagram for $\chi_l\cs$ at any loop order
 can be represented by a series of
 ``ladder segments'' separated by interactions. By  ``ladder segment'' we mean  the product
 $G_{p+q/2} G_{p-q/2}$ with vanishingly small but still finite $\q$.
 Each
    ladder segment
   contains
   integration over both high-
    and low-energy
   states.
   We define
   a
   high-energy contribution as the one where the
   $|\epsilon_\p|$
    is
    larger than $v_F|\q|$, such that the
    the poles
    of
     $G_{p+q/2}$ and $G_{p-q/2}$ are located on the same side of the real frequency axis.  This contribution can be evaluated right at $\q=0$.
     A
     low-energy contribution is  the one where
       $|\epsilon_\p|$
      is
    smaller than $v_F|\q|$, and the poles
    of
     $G_{p+q/2}$ and $G_{p-q/2}$   are on
     different
     sides of the real frequency axis.
    To obtain $\chi^{c(s)}_{l}$, we  re-arrange the perturbation series by assembling contributions
    from diagrams with a given number $M$ of
     low-energy contributions from ladder segments,
     and then sum up contributions from the sub-sets with different $M =0,1,2$, etc (see Refs.~\onlinecite{Eliashberg1962,Finkelstein2010,Chubukov2014a}).
This
procedure is demonstrated graphically in Fig. \ref{fig:bubble-series}.

\begin{figure}
  \centering
  \includegraphics[width=\hsize,clip,trim=0 70 0 70]{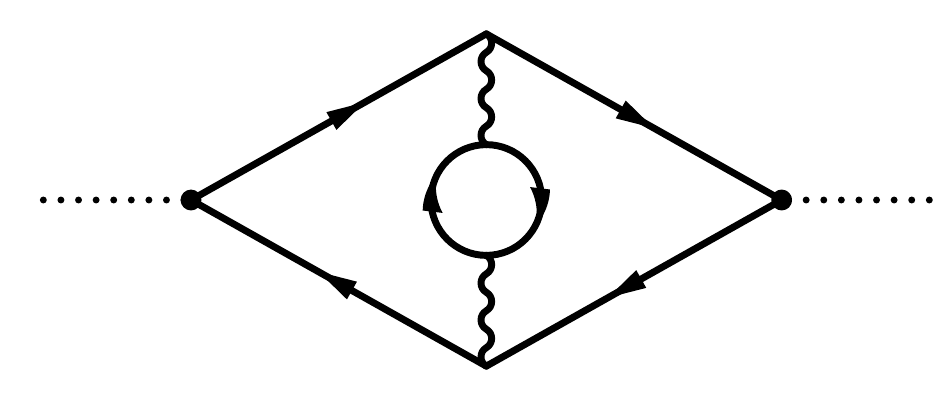}
  \caption{Example of a higher-order contribution to $\chi_l\cs$. At this order, the static interaction acquires dynamics due to
  screening
  by particle-hole pairs.
    The diagram contributes
    to the $M=0$ sector
    if
    both
    Green's functions
    adjacent
    to one of the
   external  vertices are evaluated away from the FS;
    to the $M=1$ sector,
    if one
    of them
    is evaluated on the FS and
    another one
     away from it; and to the $M=2$ sector,
     if  both are evaluated on the FS.}\label{fig:bubble-screening}
\end{figure}

We start with the $M=0$ sector.  The corresponding  contributions to the susceptibility contain products of $G^2_{p}$. Taken alone, each such term
would
vanish
on
 integration over frequency. The total $M=0$ contribution then vanishes to first order in $U(\q)$ because the static interaction does not affect the frequency integration. However, at second  and higher orders in $U(\q)$, the interaction gets screened by particle-hole bubbles and becomes a dynamical one. An example of the second-order susceptibility diagram  with screened interaction inserted into the bubble is shown in Fig.~\ref{fig:bubble-screening}.
This screened dynamical interaction contains a Landau damping term, which is non-analytic in both half-planes of complex frequency.
As a result, the product of $G^2_{p}$ and the dressed interaction at order $U^2$ and higher has both
the
double pole and a branch cut. A pole can be avoided by closing the integration contour in the appropriate frequency half-plane,
but the branch cut is unavoidable, and its presence renders the frequency integral finite.
Since one does not
have
 to make sure that the poles of the Green's functions are in the opposite half-planes,
relevant $\p$ are not confined to the FS,
and both $\omega$ and $\epsilon_\p$  are generally of order $E_F$ (or bandwidth).
 Fermions at such high energies
are strongly damped,
i.e., they
are incoherent
quasiparticles. By this reason, the $M=0$ contribution to $\chi^{c(s)}_{l}$ is labeled as an incoherent one, $ \chi^{c(s)}_{l, M=0} = \chi^{c(s)}_{l,\text{inc}}$

Next, we next move to the $M=1$ sector.  Here we select
a
 subset of diagrams with just one low-energy contribution
from some ladder segment.
 The sum of such diagrams can be graphically represented by the skeleton diagram in Fig.~\ref{fig:bubble-series}, labeled $M=1$.
   The
  ladder segment
  gives $\int_{\text{FS}} d^3p
  \left(\lambda\cs_l(\p)\right)^2
  G_{p+q/2} G_{p-q/2}$ at $\omega=0$ and $|\q|\to 0$,
  where $\int_{\text{FS}}$ denotes an integral taken close to the FS.
 Each of the Green's function in this integral can be replaced by its quasiparticle form, given by the first term in Eq.~\eqref{new_1},
  and the integral gives the static free-fermion susceptibility in Eq.~(\ref{n_14_0}) multiplied
  by a factor of $Z^2 m^*/m$.
  The
   side vertex,
   $\Lambda^{c(s)}_l$,
    is the sum of
    high-energy
     contributions from all other cross-sections either to the right or to the left of the one in which we select the low-energy piece.
    [We remind that  $\Lambda^{c(s)}_l$ is defined without the form-factor $\lambda_l\cs(p_F)$,  which was already  incorporated into $\chi_{l,0} (q)$.]
      In all these other cross-sections we can set $
     q=0$, i.e., replace $G_{p+q/2} G_{p -q/2}$ by $G^2_p$.
      These contributions  would vanish
      for a static
      interaction, but again become non-zero once we include dynamical screening  at order $U^2$ and higher. Similarly to the $M=0$ sector, the difference  $\Lambda_l^{c(s)} -1$ is determined by fermions with energies of order
      $E_F$
       (or bandwidth).
        Overall, the contribution to the static susceptibility from the $M=1$ sector is
\begin{equation}
  \chi^{c(s)}_{l, M=1} = \left(Z \Lambda^{c(s)}_l\right)^2 \frac{m^*}{m} \chi_{l,0}\cs.
  \label{n5}
\end{equation}

The sectors with $M=2$, $M=3\dots$ are  the subsets of diagrams with two, three ...
   low-energy
   parts from
   ladder segments.
    The contribution from the $M=2$ sector is represented by the skeleton diagram in Fig. \ref{fig:bubble-series} labeled $M=2$.
    A new element, compared to the $M=1$ is sector,
    is
      a fully dressed
       vertex $\Gamma^\omega$
     between fermions on the FS.
        One can easily verify that this
         vertex
          appears with
      a prefactor of $Z^2 (m^*/m)$, i.e.,
      an
       extra factor in the $M =2$ sector compared to $M=1$ is the product of $\chi_{l,0}$ and the corresponding component of the Landau function,
        as defined by Eq.~\eqref{fr_1}.
          Using (\ref{n5}), we then obtain
\begin{equation}
 \chi^{c(s)}_{l, M=1}+
  \chi^{c(s)}_{l, M=2} = \left(Z \Lambda^{c(s)}_l\right)^2 \frac{m^*}{m} \chi_{l,0}\cs \left(1 - F^{c(s)}_l\right)
  \label{n6}
\end{equation}
(the minus sign
in the second term
becomes evident if one compares
the number of the fermionic
 loops
  in the $M=1$ and $M=2$ sectors).   A simple bookkeeping analysis
shows that
the
contributions from sectors with larger $M$ form a geometric
series,
which
is re-summed
into $1/(1 + F^{c(s)}_l)$.  Collecting all contributions, we obtain
Eq.~\eqref{eq:full}.

The diagrammatic approach can be extended to the case when both
external momentum ${\bf q}$
and frequency $\omega$ are  finite (but still much smaller than $p_F$ and $E_F$, respectively),
while the ratio
\begin{equation}
  \label{eq:beta-def}
  \beta = \frac{\Omega}{v^*_F|\q|},
\end{equation}
 is
arbitrary
(here, $v^*_F = p_F/m^*= v_F (m/m^*)$). Because $\q$ and $\omega$ are small, we may still split momentum and frequency integrals
of
  $G_{k-q/2} G_{k+q/2}$ into
the low- and high-energy contributions.
 At the same time, the vertices
$\Lambda_l^{\cs}$ and $\Gamma^{\omega}$ can be still taken at $\q=0$ and $\omega=0$
because they are not sensitive to the order in which these two limits are taken.
  The
  decomposition
  of the
  perturbation series into
  the
  $M=0,1,2,\ldots$ sectors,
  depicted in Fig.~\ref{fig:bubble-series}, remains unchanged. However,
   the RPA susceptibility in Eq.~\eqref{eq:full}
   now has a nontrivial dependence on $\beta$
 and
\begin{equation}
  \chi^{c(s)}_{l} (q)  = \left(Z \Lambda^{c(s)}_l\right)^2   \chi^{c(s)}_{l,
  \text{RPA}
  } (\beta) + \chi^{c(s)}_{l,\text{inc}}.
  \label{n16}
\end{equation}
 The computation of $\chi_{l,
 \text{RPA}
 }^{\cs}(\beta)$
  is  more technically involved than
  that
  of static $\chi_{l,\text{RPA}}^{\cs}(0)$
  because different angular momentum channels no longer decouple.

Consider first the limit $\omega \ll v_F |\q|$. For even $l$,
 the quasiparticle contribution from
 $M=1$ sector is
\begin{eqnarray}
  && \chi^{c(s)}_{l,M =1, \text{RPA}} (q) \approx
  \frac{m^*}{m}\chi^{c(s)}_{l,0}\left(1 +
 i \alpha_l \beta\right),
\label{n12}
\end{eqnarray}
where $\alpha_l=1$ if $l=0$ and $\alpha_l=2$ if $l = 2
n$, $n >0$. For odd $l$, the expansion starts with $\beta^2$
--this means
that Landau damping is suppressed in odd momentum channels.\cite{oganesyan:2001}
For $M>1$, the contribution proportional
 to
 $\beta$ can come from any of the $M$ cross-sections, yielding a combinatorial factor of $M$. Summing up the series one finds
~\cite{Wu2018}
\begin{equation}
  \chi^{c(s)}_{l,\text{RPA}} (\q) =
   \chi_{l,0}\cs \frac{m^*}{m}
   \left(\frac{1}{1 + F^{c(s)}_l} +
    i \frac{\alpha_l \beta}{(1 + F^{c(s)}_l)^2}\right),
      \label{n15}
\end{equation}
where $l=2n$ is an even number.
For $l=0$, the equation above reproduces the known result for the quasi-static limit
of the charge and spin susceptibilities,\cite{Nozieres1999,baym:book} obtained by solving the kinetic equation for a FL.

In the opposite limit of $\beta \gg 1$,
only the $M=0,1,2$ elements of the bubble series need to be
 included,
 because
 higher order terms are small in $1/\beta$. Some further analysis then yields~\cite{Leggett1965,Wu2018}
 \begin{widetext}
  \begin{eqnarray}
    &&\chi^{c(s)}_{l=0,\text{RPA}} (\q)  =
    -\frac{1}{2}
     \beta^2 \chi^{c(s)}_{l=0,0}  \frac{m^*}{m} \left(1 +  F^{c(s)}_1\right),
     \nonumber \\
    &&\chi^{c(s)}_{l=1} (\q)  =
     -\frac{3}{4}
     \beta^2 \chi^{c(s)}_{l=1,0} \frac{m^*}{m} \left(1 +
       \frac{2}{3} F^{c(s)}_0 + \frac{1}{3} F^{c(s)}_2\right),
         \nonumber \\
    &&\chi^{c(s)}_{l>1} (\q)  =
    -\frac{1}{2}
    \beta^2 \chi^{c(s)}_{l,0}
     \frac{m^*}{m} \left(1 + \frac{1}{2} \left( F^{c(s)}_{l-1} + F^{c(s)}_{l+1}\right)\right),
        \label{n11}
  \end{eqnarray}
\end{widetext}
where $\chi^{c(s)}_{l,0}$ is the free-fermion static susceptibility in the corresponding channel.

For generic $\beta$,
the  form of
$\chi^{c(s)}_l (q)$ is rather involved for all $l$, including $l=0$. We illustrate
the behavior of $\chi^{c(s)}_l (q)$
 for the simplest case of the
 charge and spin susceptibilities
 ($l=0$ and $f_0(|\p|) = 1$).  Analyzing the
 series of bubbles, we find
 \beq
  \chi^{c(s)}_{l=0,\text{RPA}} (q) = \frac{m^*}{\pi}  \frac{{\bar \chi} (q)}{1 + F^{c(s)}_{l=0} {\bar \chi} (q)},
  \label{c1}
\end{equation}
where ${\bar \chi} (q)$ is given by
\begin{equation}
  {\bar \chi} (q) = K_0 - 2 \sum_{n,m >0} F^{c(s)}_{n} K_n K_m S^m_n,
  \label{c4}
\end{equation}
\begin{align}
    K_n (q) &= -\int \frac{d \theta}{2\pi} \cos{n \theta} \frac{v^*_F |\q| \cos {\theta}}{\Omega - v^*_F |\q| \cos{\theta} + i \delta_{\Omega}} \nonumber \\
              &= \delta_{n,0} - \frac{\beta}{\sqrt{(\beta)^2-1+i\delta}}(\beta-\sqrt{(\beta)^2-1})^{|n|}
  \end{align}
and
$S^m_n$ is a solution of
the linear system
\begin{equation}
  S^m_n + \sum_{m_1 >0} Q_{n,m_1} F^{c(s)}_{m_1} S^m_{m_1} = \delta_{n,m}
  \label{c5}
\end{equation}
in which $Q_{n,m} = K_{n+m} + K_{n-m}$.
In the static limit $K_0 =1$
and  $K_{n>0} =0$. Then ${\bar \chi} (
q) =1$, and Eq. (\ref{c1}) reduces to Eq. (\ref{eq:chi-c-s-mom}) for the static susceptibility.
As an additional simplification,
we consider the case when
 all Landau parameters with $l\geq 2$ can be neglected compared to $F\cs_0$ and $F\cs_1$.
  In
 that
 case,
 the
 infinite set of linear equations in \eqref{c5} is reduced to a $2\times 2$ system.
  After some further analysis, we obtain~\cite{MaslovBoltzmann,Wu2018}
\begin{equation}
  \label{n18}
  \chi^{c(s)}_{l=0,\text{RPA}} (q) = \frac{m^*}{\pi} \frac{K_0  - \frac{2 F^{c(s)}_1 K^2_1}{1  +  F^{c(s)}_1 \left(K_0 + K_2 \right)}}{1 + F^{c(s)}_0 K_0 - \frac{2 F^{c(s)}_0 F^{c(s)}_1 K^2_1}{1  +  F^{c(s)}_1 \left(K_0+ K_2\right)}}.
\end{equation}
 We stress that
 the analysis above concerns
 only the quasiparticle
 (RPA)
 contribution,
   $\chi_{l,\text{RPA}}^{\cs}$. The
   incoherent part of
   $\chi^{c(s)}_l$ does not depend on $\beta$, as long as $|\q|\ll p_F$ and $\omega\ll E_F$.

\section{Perturbation theory}
In this  section, we  use
the
perturbation theory to verify
 the  results
 derived from the conservation laws in
  the
  previous sections.
 Specifically, we compute $\Gamma^\omega$, $Z$,  $m^*/m$, and $\Lambda_{J_c(s)}$ to second order in the interaction $U(\q)$.
 The purpose of this calculation is to demonstrate
how the interplay between the contributions from high and low energies works
 both for
 $m^*/m$ and
 $\Lambda_l^{c(s)}$.

In the earlier days of the FL theory, perturbative calculations were used as a check of
 general
 FL relations~\cite{Galitskii1957}. However, several subtle issues, e.g., whether in a direct perturbative calculation
 mass renormalization
  comes solely from low energies even in the Galilean-invariant case,
 as in Eq.~(\ref{fr_2_gi}),
were not verified till fairly recently.

\subsection{The vertex function $\Gamma^\protect\omega$}

\begin{figure*}
  \begin{minipage}{0.48\hsize}
\centering
\includegraphics[width=\hsize,clip,trim=0 0 0 0]{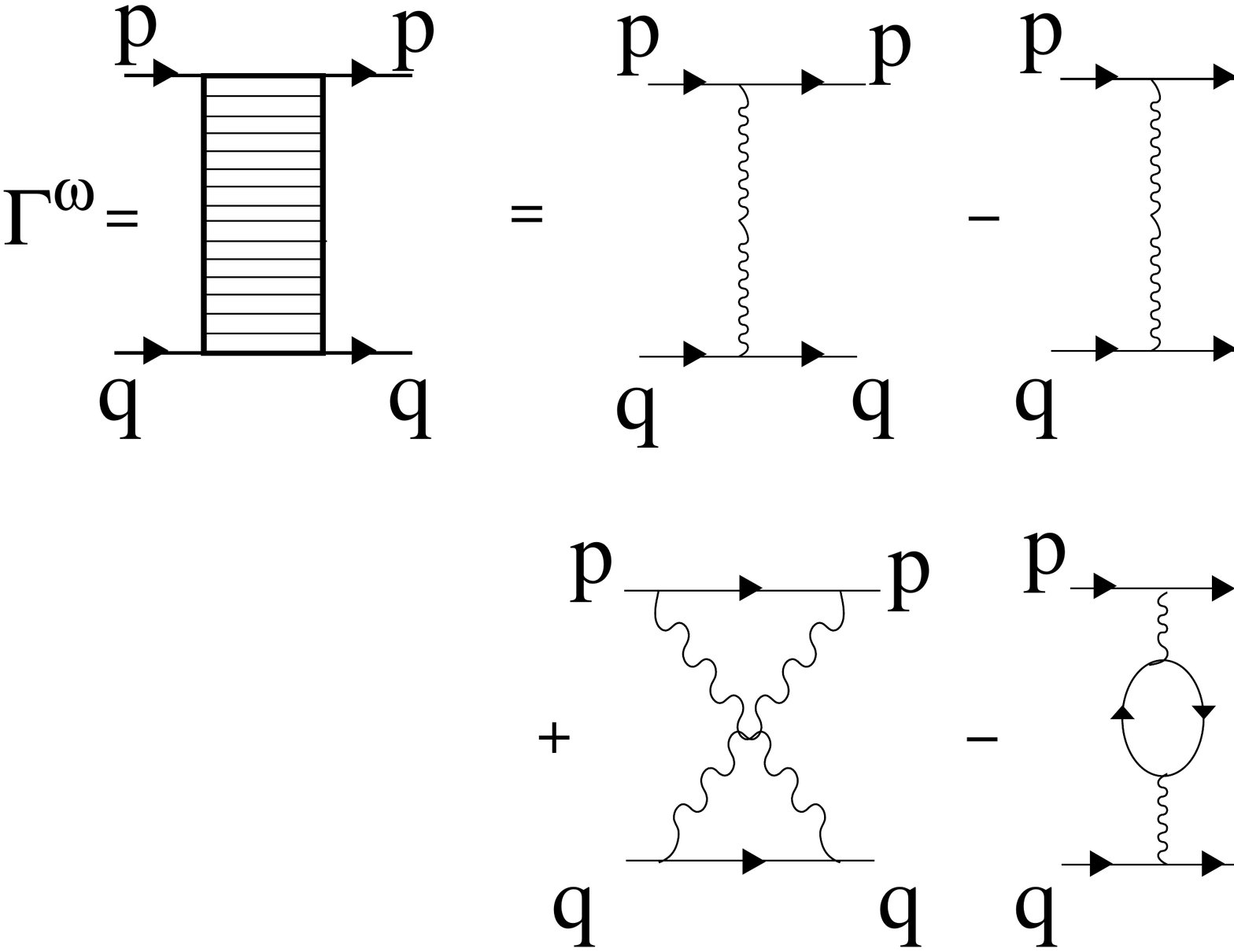}
\caption{ First and second order diagrams for the Fermi-liquid vertex $%
\Gamma^{\protect\omega}_{\protect\alpha\protect\beta,\protect\gamma\protect%
\delta}(p,q)$. The initial four-momenta $p$ and $q$ are associated with spin
projections $\protect\alpha$ and $\protect\beta$, respectively. The final
four-momenta $p$ and $q$ are associated with spin projections $\protect\gamma
$ and $\protect\delta$, respectively.
 Reproduced from Ref.~\onlinecite{Chubukov}.}
\label{fig:fig1}
\end{minipage}
\hfill
\begin{minipage}{0.48\hsize}
\centering
\includegraphics[width=0.95\hsize,clip, trim=0 300 0 0]{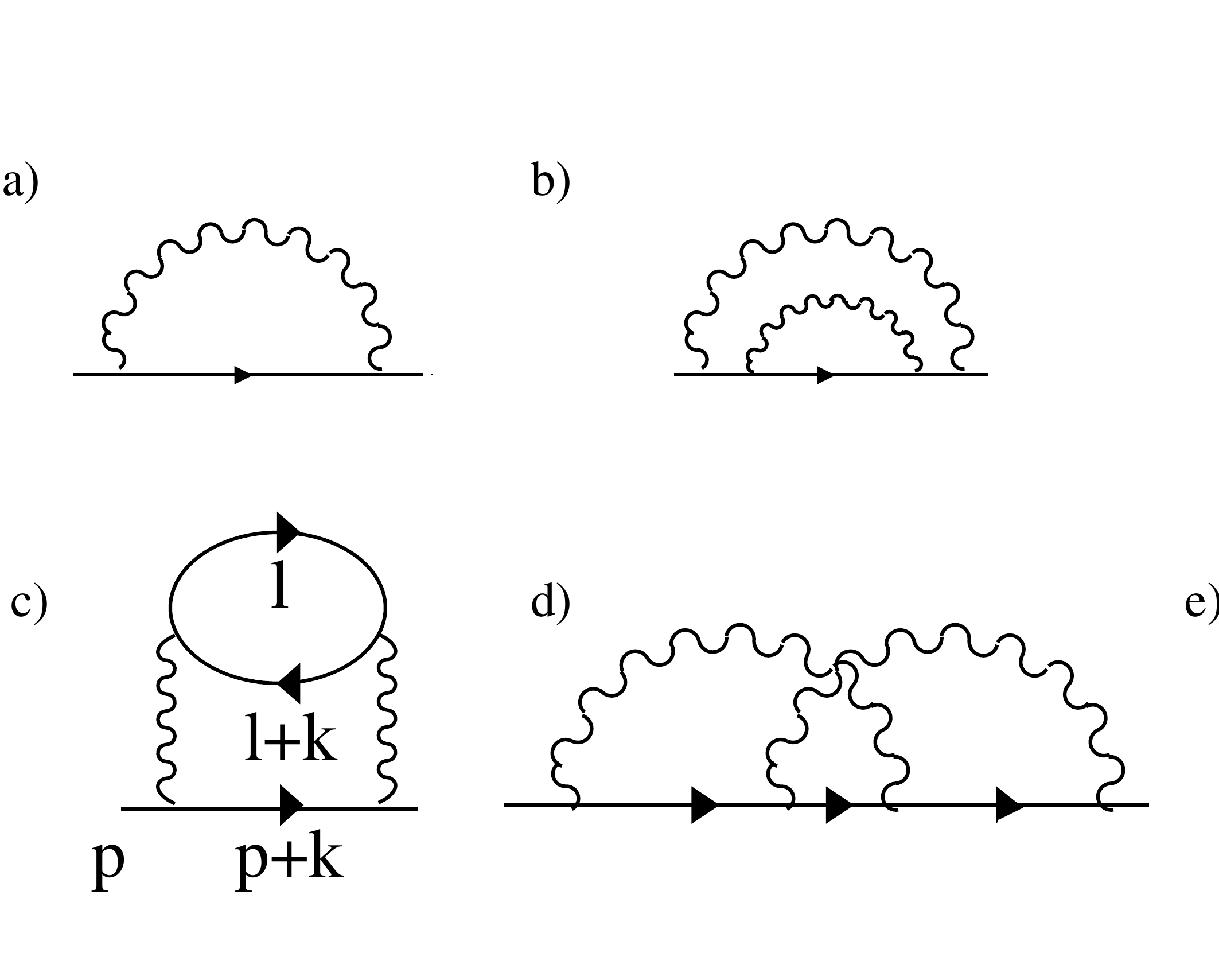}
\caption{ First- and second-order diagrams for the fermionic self-energy $%
\Sigma_{\mathrm{pert}} (\protect\omega, \protect\epsilon_{\mathbf{p}})$. For
a momentum-independent interaction $U(|{\bf q}|) =U$, only second-order diagrams renormalize the
mass and $Z$. For a momentum-dependent interaction, mass renormalization starts already at the first order,
while renormalization of $Z$ still starts at the second order.
  Diagram {\it e} is the same as  {\it c}, except for internal fermions are combined into a particle-particle rather than particle-hole pair. Reproduced from Ref.~\onlinecite{Chubukov}.}
\label{fig:fig2}
\end{minipage}
\end{figure*}

 Diagrams for $\Gamma^\omega$ to second order in $U(|{\bf k}|)$ are presented
in Fig.~\ref{fig:fig1}. The most frequently studied case is of
 the
 Hubbard (contact) interaction: $U(|{\bf k}|) = \mathrm{const}\equiv U$.
 In this case we have
\begin{widetext}
\begin{equation}
\Gamma^\omega_{\alpha\beta,
\gamma\delta}  (p_F,q) =
\delta_{\alpha\gamma}\delta_{\beta\delta}
\left[U + i U^2 \int \left(G_l G_{q-p_F+l} + G_l G_{q+p_F-l}\right)
  \frac{d^{D+1} l}{(2\pi)^{D+1}}\right] -
 \delta_
 {\alpha\delta}\delta_{\beta\gamma}
 \left[U + i U^2  \int  G_l G_{q+p_F-l}
  \frac{d^{D+1} l}{(2\pi)^{D+1}}\right].
\label{3}
\end{equation}
\end{widetext}
 Here $G_k$ stands for a free-fermion Green's function, Eq.~\eqref{new_1_1}, and $p_F$ stands for a $D+1$-momentum with zero frequency
and the spatial part equal in magnitude to the Fermi momentum and directed along $\p$.
The first term  in Eq.~(\ref{3}) is the renormalized interaction with zero
momentum transfer, the second term is obtained by antisymmetrization.  We
see that the first (``direct'') term  contains contributions from both the
particle-hole and particle-particle channels, while the second
(``exchange'') term  contains only a contribution from the particle-particle
channel.
Using the relation
\begin{equation}
\delta_
 {\alpha\delta}\delta_{\beta\gamma} = \frac{1}{2} \left( \delta_{\alpha\gamma}\delta_{\beta\delta}
   +  \boldsymbol{\sigma}_{\alpha\gamma} \cdot\boldsymbol{\sigma}_{\beta\delta}\right),
\end{equation}
we re-write Eq. (\ref{3}) as the sum of
 the
 density (charge) and spin
parts:
 \begin{equation}
\Gamma^\omega_{\alpha\beta,
\gamma\delta}  (p_F,q) =
\delta_{\alpha\gamma}\delta_{\beta\delta} \Gamma^c + \boldsymbol{\sigma}_{\alpha\gamma}\cdot\boldsymbol{\sigma}_{\beta\delta} \Gamma^s
\label{3_wed_1}
\end{equation}
with
\bea
&&\Gamma^c = \frac{U}{2} + i U^2 \int \left(G_l G_{q-p_F+l} + \frac{1}{2} G_l G_{q+p_F-l}\right) \frac{d^{D+1} l}{(2\pi)^{D+1}}, \nonumber \\
&&\Gamma^s = - \frac{U}{2} - \frac{i U^2}{2}  \int  G_l G_{q+p_F-l}
  \frac{d^{D+1} l}{(2\pi)^{D+1}}.
\label{3_wed}
\eea
 Substituting  $\sum_{\alpha\beta} \Gamma^\omega_{\alpha\beta,
\alpha\beta} (p_F,q) = 4 \Gamma^c (p_F,q)$ into the
 FL
 form of the
self-energy, Eq.~(\ref{15_a}),
 we obtain to order $U^2$
 \begin{widetext}
\begin{align}
 \Sigma_{\mathrm{FL}} (\omega, \epsilon_{\mathbf{p}}) &= \left(\omega -\epsilon_{\mathbf{p}}\right)
 U^2 \int \left(2 G_l G_{q-p_F+l} + G_l G_{q+p_F-l}\right)  G^2_q d_{ql} \nonumber \\
 &\qquad+ \epsilon_{\mathbf{p}}  \left[U^2 \int \left(2 G_l G_{q-p_F+l} + G_l G_{q+p_F-l}\right) \left(1 - \frac{{\bf p}_F \cdot{\bf q}}{p^2_F} \frac{\partial \epsilon_\q}{\partial \epsilon^{\text{par}}_{\q}}\right) G^2_q d_{ql}
 \right.
  \nn\\
 &\qquad-
 \left.
U^2 \int \left(2 G_l G_{q-p_F+l} + G_l G_{q+p_F-l}\right) \frac{{\bf p}_F \cdot{\bf q}}{p^2_F}
 \delta G^2_q  d_{ql}
 \right],
\label{15_wed}
\end{align}
\end{widetext}
where we labeled $d_{ql}
\equiv d^{D+1} q d^{D+1} l/(2\pi)^{2(D+1)}$.  The $O(U)$ term in $\Gamma^\omega$
 gives only a constant term in the self-energy
 (a shift of the chemical potential), which is omitted in the equation above.

\subsection{Direct perturbative calculation of the self-energy}
\label{pert}

We now compare Eq.~(\ref{15_wed}) with the self-energy obtained in
 the
  diagrammatic  perturbation
theory.
 As we just said, the
  term of order $U$ does not depend on $\omega$ and $\epsilon_\p$ and is therefore irrelevant  for our purposes. We  focus on
 the $U^2$ terms.  The
second-order self-energy  diagrams are shown in panels {\it b}, {\it c}, and {\it d} of Fig.~\ref{fig:fig2}.
 Only diagrams \textit{c} and \textit{d} give
 rise to $\omega$- and $\epsilon_\p$-dependent terms
 in
 the self-energy,
  the 
 diagram
   \textit{b}
 just
 adds another constant term
  to the chemical potential.
 Relabeling the fermionic momenta for $U=\text{const}$,
it is easy to see
 that
   diagram \textit{d} is equal to $-1/2$ of
 diagram \textit{c}, so we only need to consider
diagram \textit{c}.
 This diagram contains three Green's functions, two of which share a common
internal momentum.  Labeling the momenta as shown in this diagram
 and
integrating over the internal
 {D+1}-momentum $l$, we
 express
  $\Sigma_{\text{pert}}$ via
  a particle-hole
bubble.

Subtracting  from
$\Sigma_
 {\text{pert}}
 (\omega,\epsilon_{%
\mathbf{p}})$  its value at $\omega,
 \epsilon_{\mathbf{p}} =0$, we find
\begin{equation}
\Sigma_{\mathrm{pert}} (\omega, \epsilon_{\mathbf{p}}) - \Sigma (0,0) = -U^2
\int G_l G_{k-p_F+l}~ \left(G_{k+\epsilon}-G_k\right) d_{kl},  \label{n_5}
\end{equation}
where
\begin{equation}
\epsilon= \left(\omega, \epsilon_\p
\left(\frac{m \mathbf{p}_F}{p^2_F}\right)
\right)  \label{eps}
\end{equation}
 parameterizes the (small) external
 $D+1$-momentum.
 
The same self-energy can also be computed in a different way, by combing internal fermions into
 a
 particle-hole bubble.
 Re-labeling the momenta in the diagram \emph{c} as shown in panel  \emph{e} of Fig.~\ref{fig:fig2},
  we
obtain
\begin{equation}
{\tilde \Sigma}_{\mathrm{pert}} (\omega, \epsilon_{\mathbf{p}}) - {\tilde
\Sigma} (0,0) = -U^2 \int G_l G_{k+p_F-l}~ \left(G_{k-\epsilon}-G_k\right)
d_{kl}.  \label{n_55}
\end{equation}
We denote the self-energy obtained in this way as ${\tilde \Sigma}_{\mathrm{%
pert}}$
just
to distinguish it from the self-energy $\Sigma_{\mathrm{pert}} (\omega, \epsilon_{\mathbf{p}})$ in the particle-hole form.

 Since
 the two expressions for the self-energy
 must be equal,
 the Green's functions must satisfy the following identity
\begin{eqnarray}
&& \int G_l G_{k-p_F+l} \left(G_{k+\epsilon}-G_k\right) d_{kl} \notag \\
&&=\int G_l G_{k+p_F-l} \left(G_{k-\epsilon}-G_k\right) d_{kl}.  \label{9}
\end{eqnarray}
 Indeed, this identity can be proven explicitly by relabeling the fermionic momenta.\cite{Chubukov}
 To first order in $\epsilon$,
 the difference $
G_{k+\epsilon}-G_k
$ in the first line of \eqref{9} can be replaced by
 $G_{k}-G_{k-\epsilon} + O(\epsilon^2)$.
 To order $\epsilon$, therefore, identity \eqref{9} can be written as
\begin{equation}
\int \left(G_l G_{k-p_F+l} + G_l G_{k+p_F-l}\right)
\left(G_{k-\epsilon}-G_k\right) d_{kl} =0.  \label{10}
\end{equation}
Multiplying  Eq.~(\ref{10}) by $2U^2$ and
adding
 the result
 to ${\tilde \Sigma}_{\text{pert}} (\omega, \epsilon_\p)$ in  Eq.~(\ref{n_55}), we obtain
\bea
&&{\tilde \Sigma}_{\text{pert}} (\omega, \epsilon_\p) - {\tilde \Sigma}_{\text{pert}} (0, 0)\label{wed24} \\
&&= U^2 \int \left(2 G_l G_{k-p_F+l} + G_l G_{q+p_F-l}\right) \left(G_{k-\epsilon}-G_k\right) d_{kl}. \nonumber
\eea
To first order in $\epsilon$, the difference
$G_{k+\epsilon}-G_k$
 can be represented as
\bea
G_{k+\epsilon}-G_k &=& - \left(\omega - \epsilon_{\mathbf{p}}\frac{{\bf p}_F {\bf k}}{p^2_F} \frac{\partial \epsilon_\q}{\partial \epsilon^{\text{par}}_{\q}} \right)
 G^2_k  \nonumber \\
&& + \epsilon_{\mathbf{p}}\frac{{\bf p}_F {\bf k}}{p^2_F} \delta G^2_k.
\label{wed26}
\eea
The first (second) term in the equation above is a high (low)-energy contribution.
 Substituting \eqref{wed26} into (\ref{wed24}) and comparing the result with (\ref{15_wed}),
we see that ${\tilde \Sigma}_{\text{pert}} (\omega, \epsilon_\p)$ becomes equivalent to $\Sigma_{%
\mathrm{FL}}$.
 This means
 that the expressions for $m^*/m$ and $Z$, obtained from the self-energy to order $U^2$, are exactly  the
same as in the FL theory.  Using the same trick, one can also
 show that $\Sigma_{\text{pert}} (\omega, \epsilon_\p)$
 and $\Sigma_{\mathrm{FL}}$ are identical.

\subsection{Momentum-dependent interaction}

 The results for the self-energy can be readily extended to the case of a
momentum-dependent interaction.  The vertex function to order $U^2$ is
\begin{widetext}
\bea
&&\Gamma^\omega_{\alpha\beta,\gamma\delta}  (p_F,q) =
 \delta_{\alpha \gamma} \delta_{\beta \delta} \left[U(0) +i \int    \frac{d^{D+1} l}{(2\pi)^{D+1}}
U^2 (
|\mathbf{p}_F-\mathbf{l}|)\left(G_l G_{q-p_F+l} + G_l G_{q+p_F-l}\right)\right] -
\delta_{\alpha \delta} \delta_{\beta \gamma}
   \nonumber \\
&& \times \left[U(|\mathbf{q}-\mathbf{p}_F|) -i\int  \frac{d^{D+1} l}{(2\pi)^{D+1}}\left[ \left( 2 U(|\mathbf{q}-\mathbf{p}_F|) -  2 U(|\mathbf{q}-\mathbf{p}_F|) %
U(|\mathbf{p}_F-\mathbf{l}|)\right) G_l G_{l+q-p_F} -
  U(|\mathbf{p}_F-\mathbf{l}|) U(|\mathbf{l}-\mathbf{q}|) G_l G_{q+p_F-l} \right]\right] \nonumber
\label{12}
\eea
\end{widetext}
 Mass renormalization now occurs
already at the first order in $U (
|\q|)$. To this order,  perturbative and FL self-energies just coincide.
 Renormalization of $Z$ still comes from second-order diagrams.
 To second order in $U$, perturbative self-energy becomes equivalent to $\Sigma_{FL}$
with the help
 of an analog
 to (\ref{10}):
 \begin{widetext}
\begin{equation}
\int  d_{ql}U^2 (|\mathbf{p}_F -\mathbf{l}|) \left(G_l G_{k-p_F+l} +  G_l G_{q+p_F-l}\right)
\left(G_{k+\epsilon}-G_k\right).
\label{10_a}
\end{equation}
\end{widetext}
 The rest of the calculations proceeds in the same way as for the case of constant $U$.

\subsection{Where does
mass renormalization
come from
in the perturbation theory?}
\label{recons}

The issue we consider
 in this section is the separation of the perturbative self-energy into the low- and high-energy contributions.
 We note in this regard that
 (\ref{10})
 establishes a relation between these two contributions.
 Indeed,
 extracting the linear-in-$\omega$ and $\epsilon_\p$ terms from (\ref{10}) and using (\ref{wed26}), we find that Eq.~(\ref{10}) is equivalent to two equations
 \begin{subequations}
\begin{widetext}
\bea
&& \int d_{kl} \left(G_l G_{k-p_F+l} +  G_l G_{k+p_F-l}\right)
 G^2_k
 =0, \label{11_c} \\
 && \int d_{kl} \left(G_l G_{k-p_F+l} +  G_l G_{k+p_F-l}\right)
 G^2_k
  \frac{ {\bf p}_F \cdot{\bf k}}{p^2_F}
 =- \int  d_{kl}\left(G_l G_{k-p_F+l} +  G_l G_{k+p_F-l}\right) \delta G^2_k
 \frac{{\bf p}_F\cdot {\bf k} }{p^2_F}.
\label{11_b}
\eea
\end{widetext}
\end{subequations}
 Equation (\ref{11_c}) shows that a certain integral over high-energy
 states vanishes, while Eq.~(\ref{11_b}) shows that
  another integral over high-energy
  states can
  be
   expressed as
  an integral over the FS. (We remind that $\delta G_k^2$ is a projector on the FS, see \eqref{n_2}.)

 Using (\ref{11_b})
 and
  adding
  identity (\ref{10}) to either $\Sigma_{\text{pert}} (\omega, \epsilon_\p)$ or ${\tilde \Sigma}_{\text{pert}} (\omega, \epsilon_\p)$, we
  can
 redistribute
  the weights of low- and high-energy contributions
 in
 the final result.
  This
  implies
  that the same result for
 mass renormalization, computed
 either
 from $\Sigma_{\text{pert}} (\omega, \epsilon_\p)$ or ${\tilde \Sigma}_{\text{pert}} (\omega, \epsilon_\p)$,
  does not have to come from the same states.

This observation is most relevant
to a Galilean-invariant FL, where
  the phenomenological FL
 theory shows that mass renormalization comes solely from low-energy fermions.
  We
 argue
 that this is not the case if we extract $m^*/m$ from either  $\Sigma_{\text{pert}} (\omega, \epsilon_\p)$ or ${\tilde \Sigma}_{\text{pert}} (\omega, \epsilon_\p)$.

Below we present the results of the calculations separately for $D=3$ and $D=2$.

\subsubsection{3D Galilean-invariant FL}

In 3D,
 explicit expressions for $\Gamma^c$ and $\Gamma^s$
  for fermions on the FS, i.e, for $|\p|=|\q|=p_F$, and  to second order in $U$ read\cite{Abrikosov1958,agd}
\bea
&&\Gamma^c (\theta) = \frac{U}{2} + \frac{mU^2 p_F}{4\pi^2} \left(2 + \frac{\cos{\theta}}{2 \sin{\theta/2}} \log{\frac{1+ \sin{\theta/2}}{1 - \sin{\theta/2}}}\right) + ... \nonumber \\
&&\Gamma^s (\theta) = - \frac{U}{2} - \frac{mU^2 p_F}{4\pi^2}  \left(1- \frac{\sin{\theta/2}}{2} \log{\frac{1+ \sin{\theta/2}}{1 - \sin{\theta/2}}}\right) + ... \nonumber \\
&&
\label{gamma3d}
\eea
where $\theta$ is the angle between ${\bf p}_F$ and ${\bf q}_F$ and dots stand for the angle-independent $U^2$ terms.

Substituting $\Gamma^\omega$ from Eq.~(\ref{3}) into Eqs.~(\ref{2_a}) and (%
\ref{fr_2}), and evaluating the integrals,
 we obtain~\cite{Abrikosov1958,Galitskii1957}
 \begin{equation}
\frac{m^*}{m} = 1 + \left(\frac{8}{15}\right) \left(7 \ln 2 -1 \right) \left(%
\frac{mUp_F}{4\pi^2}\right)^2 \label{4_a}
\end{equation}
and
\begin{equation}
Z = 1 - 8 \ln 2 \left(\frac{mU p_F}{4\pi^2}\right)^2.  \label{4_aa}
\end{equation}

We now turn to the perturbative self-energy
 in the particle-hole representation,
  $\Sigma_{\text{pert}} (\omega, \epsilon_\p)$.
Using (\ref{wed26}) we split $\Sigma_{\text{pert}} (\omega, \epsilon_\p)$ into two parts as
\begin{equation}
\Sigma_{\mathrm{pert}} (\omega, \epsilon_{\mathbf{p}}) - \Sigma (0,0) =
\delta\Sigma_1 (\omega,\epsilon_{\mathbf{p}}) + \delta \Sigma_2
(\omega,\epsilon_{\mathbf{p}}),  \label{5}
\end{equation}
where
\begin{subequations}
\begin{eqnarray}
\delta \Sigma_1 (\omega,\epsilon_{\mathbf{p}}) &=& U^2 \int G_l G_{k-p_F+l}
G^2_k \left(\omega -\epsilon_{\mathbf{p}} \frac{\mathbf{p}_F \cdot\mathbf{k}%
}{p^2_F} \right)  d_{lk},\nn\\
\label{6_a}\\
\delta\Sigma_2 (\omega,\epsilon_{\mathbf{p}}) &=& - U^2 \epsilon_\p  \int \frac{{\bf p}_F\cdot {\bf k}}{p^2_F} \delta G^2_k
G_{k-p_F+l}
G_l
 d_{lk}.\nn\\
\label{6_b}
\end{eqnarray}
\end{subequations}
The first (second) term
 in \eqref{5}
 is a
high (low)-energy contribution.
 The
 low-energy contribution
   cannot be obtained by expanding
$\Sigma_{\mathrm{pert}}$ in
$\epsilon_\p$
 before doing the integrals.

Evaluating
the
integrals in Eqs.~(\ref{6_a}) and (\ref{6_b}), we
find
 \begin{subequations}
\begin{eqnarray}
\delta\Sigma_1 (\omega,\epsilon_{\mathbf{p}}) &=& 8 \ln 2 \left(\frac{mUp_F}{%
4 \pi^2}\right)^2 (\omega - \epsilon_{\mathbf{p}})
 \label{7a} \\
&&+\frac{4}{3} \epsilon_{\mathbf{p}} \left(4 \ln 2 -1\right) \left(\frac{%
m U p_F}{4 \pi^2}\right)^2,  \notag \\
\delta\Sigma_2 (\omega,\epsilon_{\mathbf{p}}) &=& -\frac{4}{5} \epsilon_{%
\mathbf{p}} \left(2 \ln 2 -1\right) \left(\frac{mUp_F}{4 \pi^2}\right)^2.
\label{7b}
\end{eqnarray}
\end{subequations}
Adding up the two parts, we obtain
 \bea
\Sigma_{\mathrm{pert}} (\omega, \epsilon_{\mathbf{p}}) &-& \Sigma (0,0)
 = \left(\frac{mUp_F}{4 \pi^2}\right)^2  \label{new_a}\\
&&\times \left[8\ln 2~
(\omega - \epsilon_{\mathbf{p}}) + \frac{8}{15} \left(7 \log{2}-1\right)
 \epsilon_{\bf p}
\right].\notag
\eea
 Using  Eq.~(\ref{15_a}),
we
 find
 that
the
perturbative
self-energy indeed
 gives the same results for $m^*/m$ and $Z$, as in the FL
theory, Eqs.~(\ref{4_aa}) and (\ref{4_a}).  We
 note, however, that mass
renormalization is determined by
 the prefactor
of the
{\it total}
 $\epsilon_{\mathbf{p}}$ term,
 and, according to Eqs.~\eqref{7a} and \eqref{7b},
  this prefactor comes from both  high and low energies.
  Only the sum of the two contributions recovers the  FL
formula for $m^*/m$. On the other hand,
 renormalization of $Z$ comes only
from $\delta \Sigma_1$, i.e.,  only from high energies.

\subsubsection{2D Galilean-invariant FL}

Two-dimensional analogs of Eqs.~(\ref{gamma3d})
 are
\cite{Engelbrecht}
\bea
&&\Gamma^c = \frac{U}{2} + \frac{mU^2}{2\pi} \left(2 +  \log{\cos{\theta/2}}\right) + ... \nonumber \\
&&\Gamma^s = - \frac{U}{2} - \frac{mU^2}{2\pi} \log{\cos{\theta/2}} + ...
\label{gamma2d}
\eea

Evaluating $m^*/m$ and $Z$ with the help of
 Eqs.~(\ref{2_a}) and (\ref{fr_2}), we obtain~\cite{Engelbrecht}
\begin{equation}
\frac{m^*}{m} = 1 + \frac{1}{2} \left(\frac{mU}{2\pi}\right)^2.  \label{4-1}
\end{equation}
For fermionic $Z$,
 numerical integration yields~\cite{Chubukov}
\begin{equation}
Z \approx 1 - C\left(\frac{mU}{2\pi}\right)^2,  \label{4}
\end{equation}
where $C=0.6931...$. To
 high numerical accuracy,
 $C$ is equal to $\ln 2$.

We now turn to perturbative self-energy. We again split $\Sigma_{\text{pert}}$ into
 the
 high-
  and low-energy contributions,
$\delta \Sigma_1 (\omega,\epsilon_{\mathbf{p}})$ and $\delta \Sigma_2 (\omega,\epsilon_{\mathbf{p}})$, as in Eq. (\ref{5}).
  The particle-hole bubble for 2D fermions can be obtained analytically for any $\omega$ and $|{\bf k}|$:
\begin{widetext}
\begin{equation}
\Pi_{\mathrm{ph}}(\omega ,{\bf k})=-\frac{m}{2\pi }\left[ 1+i\frac{\sqrt{2}{\tilde \omega} }{\sqrt{{\tilde k}^{2}-{\tilde k}^{4}-{\tilde \omega} ^{2}+\sqrt{\left({\tilde k}^{2}-{\tilde k}^{4}-{\tilde\omega} ^{2}\right) ^{2}-4{\tilde\omega} ^{2}{\tilde k}^{4}}}}\right],
\label{piph}
\end{equation}
\end{widetext}
where ${\tilde\omega}=2\omega m/p_F^2$ and ${\tilde k}=|{\bf k}|/2p_F$.
To calculate $\delta \Sigma_1$,
 one needs to know the entire bubble,
 while $\delta \Sigma_2$
  is determined  by the static bubble $\Pi_{\mathrm{ph}}(0,|{\bf k}|)$. Performing the angular integral in $\delta\Sigma_1$ analytically and remaining integrals numerically, and all integrals in $\delta\Sigma_2$ analytically, we obtain
  \begin{subequations}
\begin{eqnarray}
\delta\Sigma_1 (\omega,\epsilon_{\mathbf{p}}) = &&
{\tilde C} \left(\frac{mU}{2\pi}%
\right)^2\!\!\! (\omega - \epsilon_{\mathbf{p}}) + \frac{\epsilon_{\mathbf{p}%
}}{2} \left(\frac{mU}{2 \pi}\right)^2 \\
\delta\Sigma_2 (\omega,\epsilon_{\mathbf{p}}) = &&0,  \label{8}
\end{eqnarray}
\end{subequations}
with  ${\tilde C} = 0.6931\dots$
To high numerical accuracy, ${\tilde C} =C$.

 The vanishing of $\delta\Sigma_2=0$ in 2D is due to the fact that  it is expressed via a static particle-hole bubble:
\begin{eqnarray}
\delta\Sigma_2 (\omega,\epsilon_{\mathbf{p}}) &=& - \frac{\epsilon_{\mathbf{p}} U^2}{2\pi^2 v_F}%
\int^{2p_F}_0d|\mathbf{k}| \Pi_{\mathrm{ph}}(\omega=0,|\mathbf{k}|)  \notag
\\
&&\times\frac{1-|\mathbf{k}|^2/2p_F^2}{\sqrt{1-\left(|\mathbf{k}%
|/2p_F\right)^2}}. \label{s2}
\end{eqnarray}
Because $\Pi_{\mathrm{ph}}(\omega=0,|\mathbf{k}|)$ is independent of $|\mathbf{k}|
$ for $\mathbf{k}|\leq 2p_F$, the integral over $|\mathbf{k}|$ vanishes.

Casting the result into the form of Eq.~(\ref{15}), we again reproduce the FL
results for $m^*/m$ and $Z$, Eqs.~(\ref{4-1}) and (\ref{4}). However, we see
that now $m^*/m$
  comes solely from the  high-energy part of the self-energy.

If we compute 
  the
  perturbative self-energy by combining two internal fermions into
  a
  particle-particle bubble (${\tilde \Sigma}_{\text{pert}} (\omega, \epsilon_\p)$ in our notations), and again split it into high-energy and 
 low-energy contributions, $\delta{\tilde\Sigma}_1(\omega,\epsilon_{\mathbf{p}})$ and $\delta{\tilde\Sigma}_2(\omega,\epsilon_{\mathbf{p}})$, we obtain~\cite{Chubukov}
\begin{subequations}
\bea
\delta{\tilde\Sigma}_1(\omega,\epsilon_{\mathbf{p}})&&= \bar C \left(\frac{mU}{2\pi}%
\right)^2\left(\omega - \epsilon_{\mathbf{p}}\right)+\epsilon_{\mathbf{p}%
}\left(\frac{mU}{2\pi}\right)^2,\nn\\  \label{new_n3} \\
\delta{\tilde\Sigma}_2(\omega,\epsilon_{\mathbf{p}})&&= -\frac{\epsilon_{%
\mathbf{p}}}{2}\left(\frac{mU}{2\pi}\right)^2,  \label{new_n2}
\eea
\end{subequations}
where, as before, $\bar C = 0.6931\dots$.
Comparing with Eq.~(\ref{4-1}), we see that now the low-energy
contribution to mass renormalization in the particle-particle case is finite
but opposite in sign
that to mass renormalization in the FL theory. The correct sign is reproduced once we add
the low- and
high-energy contributions.

 \subsection{Direct perturbation theory for static susceptibility}
\label{sec:direct-pert-theory}

We now show that the same subtle interplay between the low- and high contributions
 also occurs for the charge/spin spin susceptibility in the channel with angular momentum $l$,
 $\chi^{c(s)}_{l}$.

Rather than going through an exhaustive analysis, we consider a single illustrative example, namely
the $l=1$ spin channel in a Galilean-invariant system.
 Our goal is to reproduce  the relation
 $\Lambda^s_{J} Z (m^*/m) =1+F_1^s$.
 Explicitly,
 we have
\begin{subequations}
\bea
 &&  \frac{1}{Z} = 1
  -  2i \int d_k \Gamma^c (\p_F,\p) \frac{(\p_F, \p)}{p^2_F} G_k^2  \label{eq:Z-GI-FL} \\
&& \Lambda_{J}^s = 1
-2 i
  \int d_k
  \Gamma^s(\p_F, \p)
  G^2_k
  \frac{(\p_F, \p)}{p^2_F}
  \label{eq:delta-lambda-1s-FL}
\eea
\end{subequations}
 The vertex functions
$\Gamma^c$ and $\Gamma^s$ to order $U^2$ i
are given by (\ref{3_wed}).
Combining the contributions from
$Z$ and $\Gamma\cs$ we obtain, to order $U^2$,
\begin{flalign}
  \label{eq:delta-lambdaZ-u2}
  \Lambda^s_{J} Z -1=
 -2
    U^2 \int d_{kl} \left (G_l G_{k-p_F + l} + G_k G_{k+p_F-l}\right)  \nn\\
  \frac{(\p_F\cdot \p)}{p^2_F}  G^2_k
\end{flalign}

As written, the integral
 on the r.h.s. of  Eq.~\eqref{eq:delta-lambdaZ-u2} is not confined to the FS.
 However, it can be
 converted
 into a
  FS
  contribution
 using
 identity
 (\ref{11_b}), which expresses the r.h.s. of (\ref{eq:delta-lambdaZ-u2}) via the integral over $\delta G^2_k$. We then obtain
 \begin{align}
  \label{eq:delta-lambdaZ-u2a}
  &\Lambda^s_{J} Z
  - 1
  =
  2
  U^2 \int  d_{kl}\left(G_l G_{k-p_F+l} +  G_l G_{k+p_F-l}\right) \delta G^2_k.
 \frac{{\bf p}_F\cdot {\bf k} }{p^2_F}
 \end{align}
 Finally, we use Eq.~(\ref{3_wed}) and
 re-write the r.h.s. of (\ref{eq:delta-lambdaZ-u2a}) as
\begin{widetext}
\begin{equation}
\label{llla}
 2 U^2 \int  d_{kl}\left(G_l G_{k-p_F+l} +  G_l G_{k+p_F-l}\right) \delta G^2_k
 \frac{{\bf p}_F\cdot {\bf k} }{p^2_F} = 2 \int \frac{d\theta}{2\pi} \left( F^s (\theta) - F^c (\theta) \right) \cos{\theta}  = F^s_{1} - F^c_{1}.
 \end{equation}
\end{widetext}
Substituting
this
into (\ref{eq:delta-lambdaZ-u2a}), we obtain $Z \Lambda^s_{J} = (1 + F^s_{1} - F^c_{1})$, which, to order $U^2$, is equivalent to
 $Z \Lambda^s_{J} = (1 + F^s_{1})/(1 + F^c_{1})$,
 as in  Eq. \eqref{mo_3}.

We emphasize that only the product $\Lambda^s_{J} Z$ can be expressed via
an  integral over the FS.  Taken separately, $ \Lambda^s_{J}$ and $Z$ are determined by integrals which are not confined to the FS.

\section{conclusions}
In conclusion, we reviewed certain aspects of the microscopic FL theory. We argued
that this theory is based on five Ward identities,
which
follow
  from conservation laws.
The first two identities (the Pitaevskii-Landau relations\cite{Pitaevskii1960}) follow from $U(1)$ symmetry and reflect charge conservation. The next two (the Kondratenko relations\cite{kondratenko:1964,kondratenko:1965}) follow from $SU(2)$ symmetry and reflect spin conservation. The last, fifth relation,
follows from  translational symmetry and reflects momentum conservation.
This last identity
was derived originally for a Galilean-invariant system,\cite{Lifshitz1980} but is generalized here for any translationally invariant system and thus
can be attributed to momentum conservation.
 These identities
  express quasiparticle $Z$ and the effective mass $m^*$ in terms of the vertex function.
  In addition,
  they
  impose certain constraints
on the interplay between
 low- and high-energy
contributions to observable quantities.
    These constraints imply that  extra care is needed in
integrating out contributions from high-energy fermions.
 For example, the low- and high-energy contributions to the susceptibilities of charge and spin currents cancel each other, so that Pomeranchuk instabilities towards phases with spontaneously generated charge and spin currents are impossible.\cite{Leggett1965,Kiselev}
Even more so, the corresponding susceptibilities
 are not renormalized at all
 by the interaction.~\cite{Leggett1965} On the other hand,
an instability towards a phase
 with
  the order parameter,
  which has  either  the same symmetry as  the charge or spin current
  but
  a
   different
   form-factor, or
   a
   different symmetry,
   is not forbidden by
conservation laws.\cite{Wu2018}

We also demonstrated how the constrains imposed by conservation laws can be derived diagrammatically, and along the same lines, provided a diagrammatic derivation of the Leggett formula\cite{Leggett1965} for the charge and spin susceptibility in a channel with arbitrary angular momentum.
Finally, we
  illustrated the interplay between the low- and high-energy contributions by calculating the FL interaction vertex, effective mass, quasiparticle residue, and susceptibility to second-order in interaction.

\begin{acknowledgements}
We thank  J. Schmalian, P. Woelfle, and Y. Wu  for  valuable discussions. The work was supported by  NSF DMR-1523036 (A.V.C. and A.K.) and  NSF DMR-1720816 (D.L.M.).
\end{acknowledgements}

\bibliographystyle{apsrev4-1}
\bibliography{C_S_susceptibility_1}

\end{document}